\documentclass{osa-article}

\journal{osac}
\usepackage{comment}
\usepackage{subfigure}

\articletype{Research Article}
\begin{document}

\title{Secret Key Distillation over Satellite-to-satellite Free-space Optics Channel with Eavesdropper Dynamic Positioning}

\author{Ziwen Pan\authormark{*} and Ivan B. Djordjevic}

\address{Department of Electrical \& Computer Engineering, College of Engineering, 
the University of Arizona, 1230 E Speedway Blvd, 
Tucson, AZ 85719, USA\\
}

\email{\authormark{*}ziwenpan@email.arizona.edu} %% email address is required

% \homepage{http:...} %% author's URL, if desired

%%%%%%%%%%%%%%%%%%% abstract %%%%%%%%%%%%%%%%
%% [use \begin{abstract*}...\end{abstract*} if exempt from copyright]

\begin{abstract}
The conventional omnipotent eavesdropper assumption in  quantum cryptography study  can be too strict for some realistic scenarios. In this paper, we study the secret key distillation over a satellite-to-satellite free space optics channel in which we assume that the eavesdropper (Eve) is restricted in her ability of power collection due to the limited size of her aperture but can change the position of her aperture to gain advantages over the communication parties (Alice and Bob) and we determine the achievable  key rate lower and upper bounds with respect to different scenarios. We first study the case where Eve is behind Bob and we %. We investigate Eve's strategy of  positioning her aperture and 
prove that the optimal eavesdropping strategy for her in long-distance transmission case is to place her aperture on the beam transmission axis and set Bob-to-Eve distance equal to Alice-to-Bob distance. We  also show that the achievable  key rate would be characterized by a Bessel function integral related to Eve's position in a short-distance transmission case. We then investigate the case where Eve is before Bob and show similar results with Eve's and Bob's roles exchanged. %We show input power dependency without assumptions on Bob's detection scheme and then we determine lower bounds with input power optimized in different scenarios for specific discrete variable (DV) and continuous variable (CV) protocols for comparison. 
%We demonstrate that significantly higher SKR lower bounds can be achieved compared to traditional unrestricted Eve scenario. We conclude that this model is suitable for extended analysis in many light-gathering scenarios including both the near and the far-field and for different wavelengths. %\textbf{For other problems, refer to the dailydocument of 20201202.}
\end{abstract}

%%%%%%%%%%%%%%%%%%%%%%%%%%  body  %%%%%%%%%%%%%%%%%%%%%%%%%%
\section{Introduction}
Traditionally, quantum cryptography aims at providing unconditional informational security with the assumption of an omnipotent eavesdropper whose power is only limited by the laws of physics.  BB84, the first protocol for quantum key distribution (QKD), was developed in 1984 by Charles Bennett and Gilles Brassard~\cite{bennett1984quantum} where the security is based on no-cloning theorem and one-time pad encryption. This started a series of work on discrete variable QKD (DV-QKD)~\cite{inoue2002differential,hwang2003quantum,pan2017quantum,PhysRevLett.108.130503,PhysRevLett.108.130502,PhysRevA.86.062319} that uses single photon sources and detectors, which can be challenging in implementations. % transmitted, providing security while also putting forward rigorous conditions for realistic implementations, and various protocols have since been proposed. 
To improve QKD in realistic implementations, especially to be more compatible with existing optical communication networks and equipment,  continuous variable QKD (CV-QKD) has been developed, for example, protocols based on coherent laser source and heterodyne detection~\cite{LPFPP18, DL15}, with  promising experimental progress~\cite{zhang2019continuous,zhang2020long}.

However, the assumption of an omnipotent eavesdropper is not the case for realistic applications since Eve would also be limited to realistic factors other than the laws of physics. For example, in a wireless communication channel, it would be unrealistic to assume that Eve has an infinite-sized receiver aperture with unlimited power collecting ability although this restriction does not directly come from the laws of physics.  In our recent papers~\cite{pan2019secret,8849223} we presented the theoretical analysis of secret key distillation  with a restricted Eve by performing achievable  key rate calculation with restrictions on Eve's collecting ability. We then analyzed, in our follow-up works, specific scenarios where the restriction is reasonable and properly quantified, for example the limited aperture size of Eve's receiver in the same plane of Bob~\cite{pan2020secretOE,pan2020secretQ2}, and the  corresponding defense strategy such as setting an exclusion zone around the legitimate communication parties' apertures~\cite{pan2020secretExZo,pan2020securityICTON}.

In this work, as a followup of~\cite{pan2020securityICTON,pan2020secretSPPCom}, we investigate the case where Eve has a limited sized aperture that can be positioned at any location, which would be common in secret key distillation in space, eg, with a spy satellite that can move around to gain Eve advantages against the communication parties.
%In the next two sections we characterize two typical scenarios where eavesdropper's receiving aperture (Eve) is close to the legitimate communication party's (Bob). 
%In Sec.~\ref{ExZo} 
In Sec.~\ref{BehindBob} we first introduce the problem setup of the limited-sized aperture of Eve behind Bob and calculate the channel parameters assuming that Gaussian beam is transmitted. We first analyze the case where Eve's aperture is on the beam transmission axis in Sec.~\ref{D=0}, with the input power dependency of achievable key rate lower bounds shown in Sec.~\ref{InputPowerDependency} and optimized input power results shown in Sec.~\ref{LowerBoundswithOptimizedInputPower}. We also show the proof of Eve's optimal eavesdropping distance in a long-distance transmission scenario and compare the short-distance case characterization with the results from the study of the Arago Spots~\cite{harvey1984spot,reisinger2017relative}.  Then we optimize Eve's position off the beam transmission axis in Sec.~\ref{Doptimized} and show how Eve can gain advantages by moving her aperture off axis while approaching Bob. In Sec.~\ref{BeforeBob} we look into the  case where Eve's aperture is before Bob and determine the achievable key rate lower and upper bounds with respect to Eve's location between Alice and Bob. We show that in this case Eve's strategy would be easier approaching Alice than Bob.  For the above different scenarios of our study, we also showcase the secure key rate (SKR) lower and upper bounds results with respect to specific CV- and DV-QKD protocols with input power optimized under imperfect reconciliations.

\section{Eve behind Bob}\label{BehindBob}
%In this section and the next section we used z for distance. Be noticed if there are remnants that haven't been changed.

\begin{figure}[htbp]%[htbp] %% Figure 4 
\centering{\includegraphics[height=13pc]{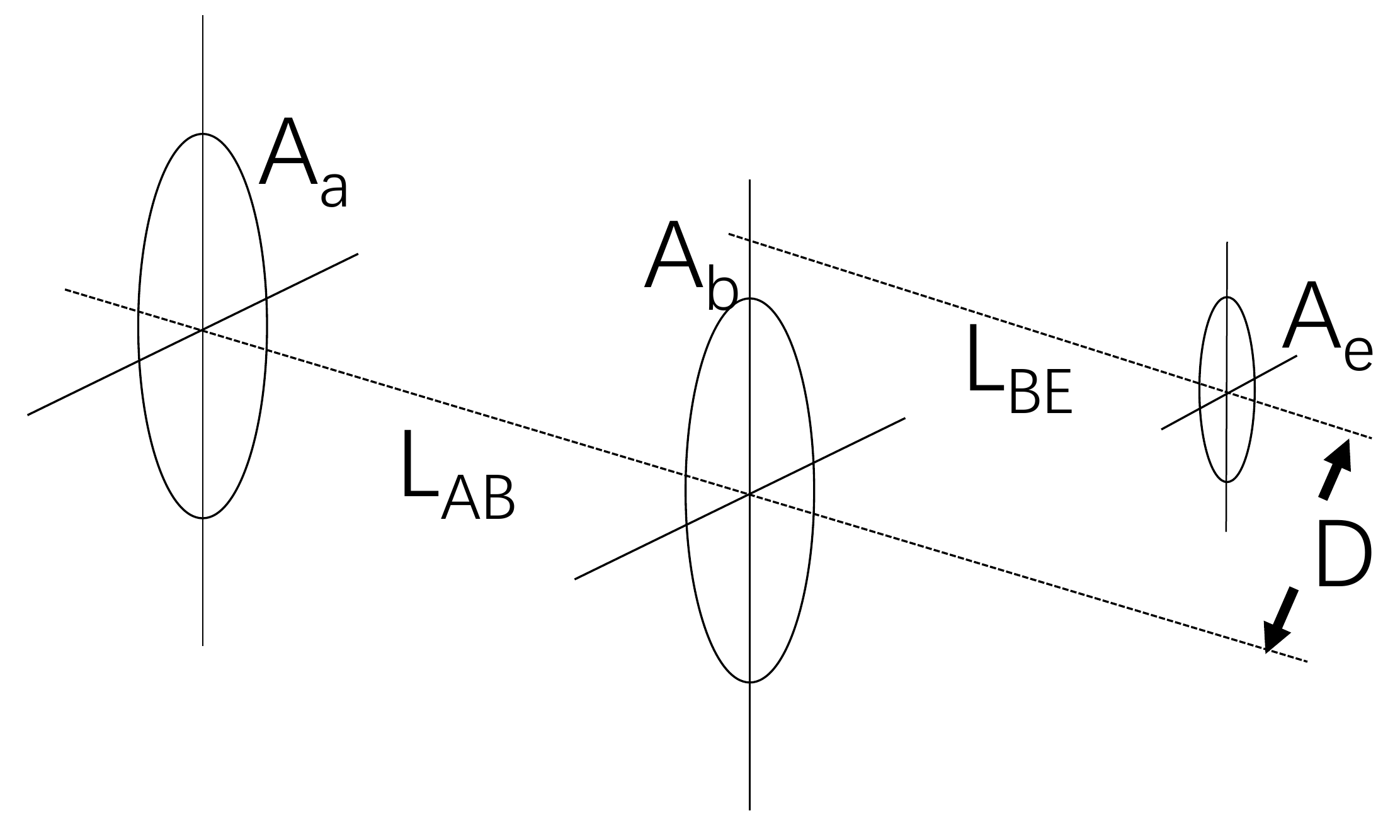}}
\caption{Geometric setup of the eavesdropper dynamic positioning scenario. Here we use $A_a$ to denote the transmitting aperture (Alice) area, $A_b$ to denote receiving  aperture (Bob) area and $A_e$ to denote eavesdropper aperture (Eve) area. %The exclusion zone area is denoted as $A_e$ which is a ring around Bob's aperture. 
$L_{AB}$ is the  transmission distance between Alice and Bob and $L_{BE}$ is the distance between Bob's aperture plane and Eve's aperture plane. $D$ is the distance between the center of Eve's aperture and the beam propagation axis. %Eve can collect photons outside of the exclusion zone.
\label{EvePo2}}
\end{figure}

\noindent In this section we first investigate the case where Eve's aperture is behind Bob. As is illustrated in Fig.~\ref{EvePo2}, we assume that the area of transmitter aperture is $A_{Alice}$ ($A_a$) with radius $r_{Alice}$ ($r_a$), the area of receiver aperture is $A_{Bob}$ ($A_b$) with radius $r_{Bob}$ ($r_b$), and the area of eavesdropper aperture is $A_{Eve}$ ($A_e$) with radius $r_{Eve}$ ($r_e$). $L_{AB}$ is the  transmission distance between Alice's aperture plane and Bob's aperture plane and $L_{BE}$ is the distance between Bob's aperture plane and Eve's aperture plane. Here $D$ denotes the distance between the center of Eve's aperture and the beam propagation axis. We also assume that Gaussian beam has been transmitted with beam waist radius $W_0$ equal to transmitter aperture radius. Since Gaussian beam is cylindrical symmetric along the propagation path, %$D$ is sufficient to describe how much power Eve can collect at her current location and thus  
we will use $D$ as Eve's position. 

\begin{comment}
\noindent In this section we introduce the Limited Aperture Scenario, in which we assume that Eve has a limited-sized aperture %with size 
$A_{e}$ (radius $r_e$), as in Fig.~\ref{EvePo2}. In this paper we only look at this straightforward case which actually gives us very interesting results when Eve's aperture is in the same plane as Bob's aperture with no overlapping. In this case, the optimal position is when Eve's aperture is tangential to Bob's aperture, %for a Gaussian wavefront, 
as illustrated in Fig.~\ref{EvePo2}. Here we denote the area of transmitter aperture as $A_a$ (Alice), the area of receiver aperture as $A_b$ (Bob) and the area of eavesdropper aperture as $A_{e}$ (Eve).  %More generalized scenarios will be included in a complete version of this paper which will be available online soon. 
\end{comment}

%As is in Fig.~\ref{EvePo2}, 

The   transmitted Gaussian beam field amplitude $U$ in space can be expressed as:

\begin{equation}
    U(r,L)=E_0\frac{W_0}{W(L)}\exp\left({-\frac{r^2}{W^2(L)}}\right)\exp\left(-i\left(kL+k\frac{r^2}{2R(L)}-\psi(L)\right)\right).
\end{equation}
where $r$ is the distance from the observation point to the beam center axis, $L$ is the propagation distance,  $E_0$ is the field amplitude at the center of the beam at its waist, $W_0$ is the waist radius of the Gaussian beam, which we set to be equal to $r_{Alice}$ throughout this work. $k$ is the wave number with
\begin{align}
W(L)&=W_0\sqrt{1+(L/z_0)^2},\\
    R(L)&=L\left(1+\left(\frac{z_0}{L}\right)^2\right),\\
    \psi(L)&=\arctan\left(\frac{L}{z_0}\right),\\
    z_0&=nW_0^2\frac{\pi}{\lambda}.
\end{align}
Here $z_0$ is the Rayleigh length, $W(L)$ is the spot size parameter, $R(L)$ is the radius of curvature and $\psi(L)$ is the Gouy phase. $n$ is the refractive index, which is equal to 1 in space.  $\lambda$ is the wavelength of the beam being transmitted, which we set to 1550nm throughout this work.

%If we assume that Eve's aperture can't have any overlap with Bob's aperture and is in the same plane with Bob's aperture, since the power distribution over receiving plane is 2 dimensional Gaussian, then Eve's optimal position to put her lens would be right next to Bob's lens, as is illustrated in Fig.~\ref{EvePo2}. And it's obvious that Eve cannot collect more light by getting behind Bob's aperture plane since the beam propagated would go through more diffraction in that way. Without loss of generality due to the symmetric power distribution on receiving plane, we can assume that Eve puts her aperture right above Bob's aperture and do the integration respectively to get each one's receiving power $P$. We also assume Bob's aperture radius to be $r_b$ and Eve's to be $r_e$.

%If we assume that Eve's aperture is located in the same plane as Bob's aperture and restricted to have no overlapping regions, the optimal position is when Eve's aperture is tangential to Bob's aperture for a Gaussian wavefront, as illustrated in Fig.~\ref{EvePo2}. 
%If Eve's aperture is of a constant size and is moved to the plane behind Bob, no additional power can be obtained. This is due to the diffraction effects induced by Bob's aperture. 
%Without loss of generality due to the symmetric power distribution on receiving plane, we assume that Eve puts her aperture right above Bob's aperture and perform the integration respectively to get each one's receiving power $P_{Bob}$ and $P_{Eve}$. We also assume Bob's aperture radius to be $r_b$ and Eve's one to be $r_e$.

Bob's received power can be easily calculated:
\begin{align}
    P_{Bob}&=\int_{0}^{2\pi} \int_{0}^{r_b} \|U(r,L_{AB})\|^2 \, rdrd\theta=2\pi  \int_{0}^{r_b} \|U(r,L_{AB})\|^2 \, rdr,\label{PBob}
\end{align}

The total power in this Gaussian beam is:
\begin{equation}
    P_{total}=E_0^2\frac{\pi  W_0^2}{2},
\end{equation}

For different scenarios we calculate Eve's received power ($P_{Eve}$) respectively, then we can calculate Alice-to-Bob transmissivity ($\eta$) and Eve's fraction of collected power ($\kappa$)~\cite{pan2019secret} by: 
\begin{align}
    \eta&=\frac{P_{Bob}}{P_{total}},\label{eta}\\
    \kappa&=\frac{P_{Eve}}{(1-\eta)P_{total}}\label{kappa},
\end{align}
Also for noise frequency dependence we use the black body radiation function:
\begin{equation}
n_e=\frac{1}{e^{\frac{hf}{kT}}-1}\label{blackradia}.
\end{equation}
where $n_e$ is the mean photon number per mode for the thermal noise, $h=6.626*10^{-34} m^2kg/s$ is the Planck constant, $k=1.38064852*10^{-23} m^2 kg/(Ks^2)$ is the Boltzmann constant, $T=3\text{K}$ is the space temperature and $f$ is the center frequency in transmission.

\subsection{$D=0$}\label{D=0} First we look at the straightforward case where $D=0$, which would be the optimal position for Eve when $L_{BE}$ is large as the cropped Gaussian beam right after Bob's aperture plane would have reconverged, leaving a major portion of its photons at the center. In this case, Eve's received power can be expressed as follows with the Rayleigh-Summerfeld transfer function:
\begin{align}
P_{Eve}&=\int_{0}^{2\pi} \int_{0}^{r_e} \|U_{Eve}\left(l,\phi\right)\|^2 \, ldld\phi,\label{PEve}\\
    U_{Eve}\left(l,\phi\right)&=\frac{L_{BE}}{i\lambda}\int_{0}^{2\pi} \int_{r_b}^{\infty} U\left(r,L_{AB}\right)\frac{e^{ikr_{12}}}{r_{12}^2} \, rdrd\theta,\label{UEve}\\
    r_{12}&=\sqrt{L_{BE}^2+l^2+r^2-2rl\cos{(\phi-\theta)}}.
\end{align}
where $r_{12}$ is the distance between a point on Eve's aperture plane (polar coordinate $(r,\theta)$) and a point on Bob's aperture plane (polar coordinate $(l,\phi)$). $U_{Eve}(l,\phi)$ is the field amplitude at $(l,\phi)$ on Eve's aperture plane.

To start with, we look at the input power dependency in this scenario with several selected Bob-to-Eve distances ($L_{BE}$).

\subsubsection{Input Power Dependency}\label{InputPowerDependency}

\noindent In this section we look into the achievable  key rate lower bounds input power dependency by applying the methods of secure key rate analysis in~\cite{pan2019secret}. %based on Eqs.~(\ref{eta}), (\ref{kappa}) and (\ref{blackradia}). 
Recall that the direct ($K_\rightarrow$) and reverse ($K_\leftarrow$) reconciliation lower bounds    in a quantum thermal noise wiretap channel with reconciliation efficiency $\beta$  are  given:
\begin{align}
    K_\rightarrow &\geq \beta g\left(n_e(1-\eta)+\eta\mu\right)-\sum_i g\left(\frac{\nu^{ER}_i-1}{2}\right)\nonumber\\
&-\beta g\left(n_e(1-\eta)\right)+g\left(n_e(1-\eta\kappa)\right)\label{LBDmu},\\
K_\leftarrow &\geq \beta g(\mu)-\sum_i g\left(\frac{\nu^{ER}_i-1}{2}\right)\nonumber\\
&-\beta g\left(\mu-\frac{\eta\mu(1+\mu)}{1+n_e-n_e\eta+\eta\mu}\right)+\sum_i g\left(\frac{\nu^{ER}_{y_i}-1}{2}\right)\label{LBRmu},\\
g(x)&=(x+1)\log_2(x+1)-x\log_2(x),\label{thentropy}
\end{align}
where $\mu$ is the average photon number that Alice transmits to Bob per input mode. %If we make reasonable assumptions on $A_a$, $A_b$ (radius $r_a$, $r_b$) and use $A_{e}$ (radius $r_{e}$) as a parameter then we can plot direct (dashed curves) and reverse (solid curves) SKR lower bounds against input power $\mu$. Here we set transmission wavelength as 1550nm. %Below in this section we use yellow curves to denote the case when Eve's aperture is larger than Bob's ($r_{e}>r_b$) whereas blue curves denote the case when Eve has a smaller aperture ($r_{e}<r_b$). 

\begin{figure}[htbp]
\centering
%\begin{minipage}[t]{0.49\textwidth}
\centering
\includegraphics[width=8.8cm]{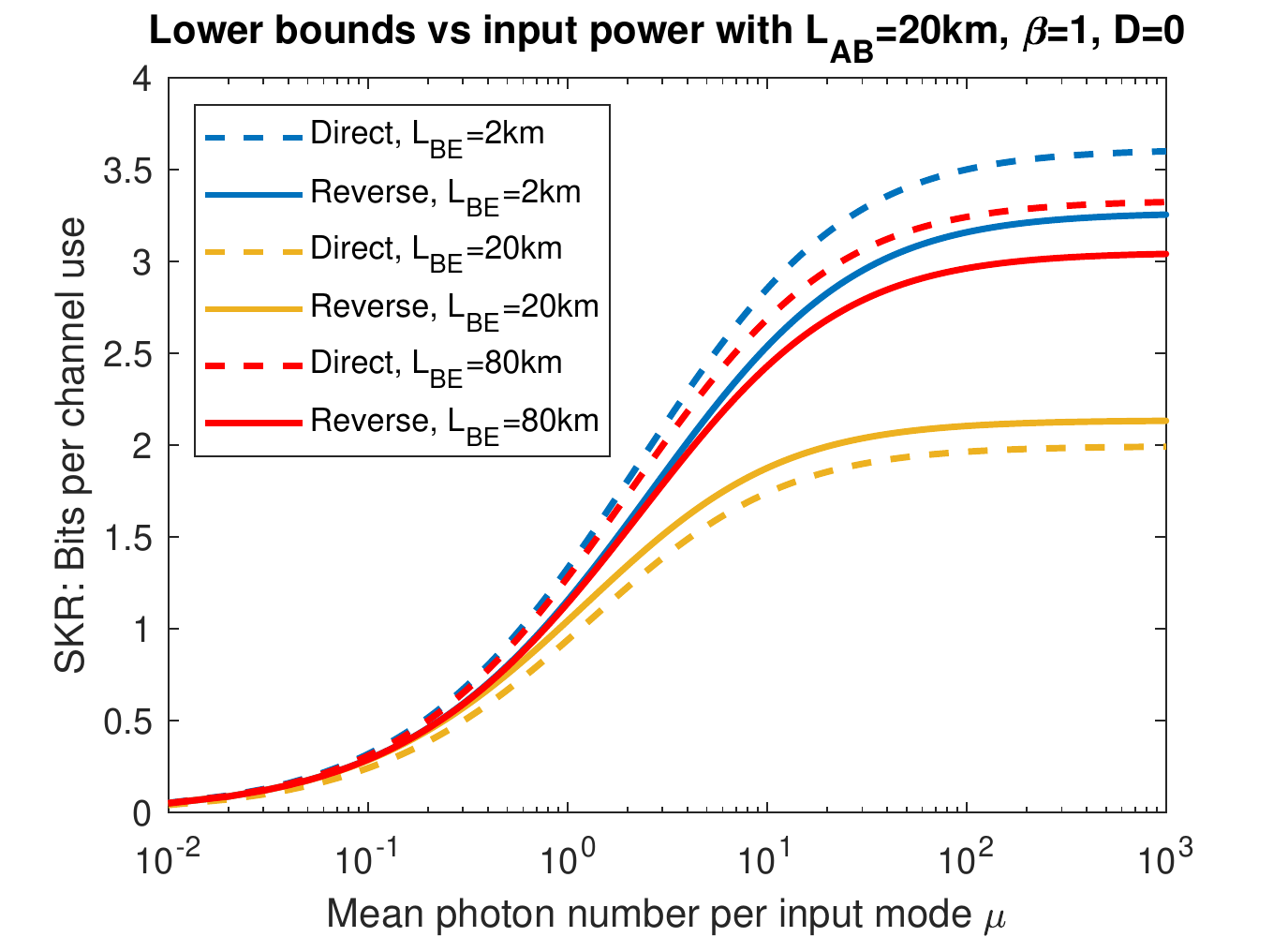}
\caption{Lower bounds vs. input power. 
The Alice-to-Bob distance $L_{AB}$ is 20km. Bob-to-Eve distances $L_{BE}$ and reconciliation schemes (direct versus reverse) are specified in the legend. Reconciliation efficiency $\beta$ is set to 1. Transmission center wavelength $\lambda$ is set to 1550nm. Transmitted Gaussian beam waist radius is set to $W_0=r_a=10$cm. Bob and Eve aperture radius are set to $r_b=r_e=10$cm. \label{Index202009151945}}%ploted by Index202009151945
\end{figure}

\begin{figure}[htbp]
\centering
%\begin{minipage}[t]{0.49\textwidth}
\centering
\includegraphics[width=8.8cm]{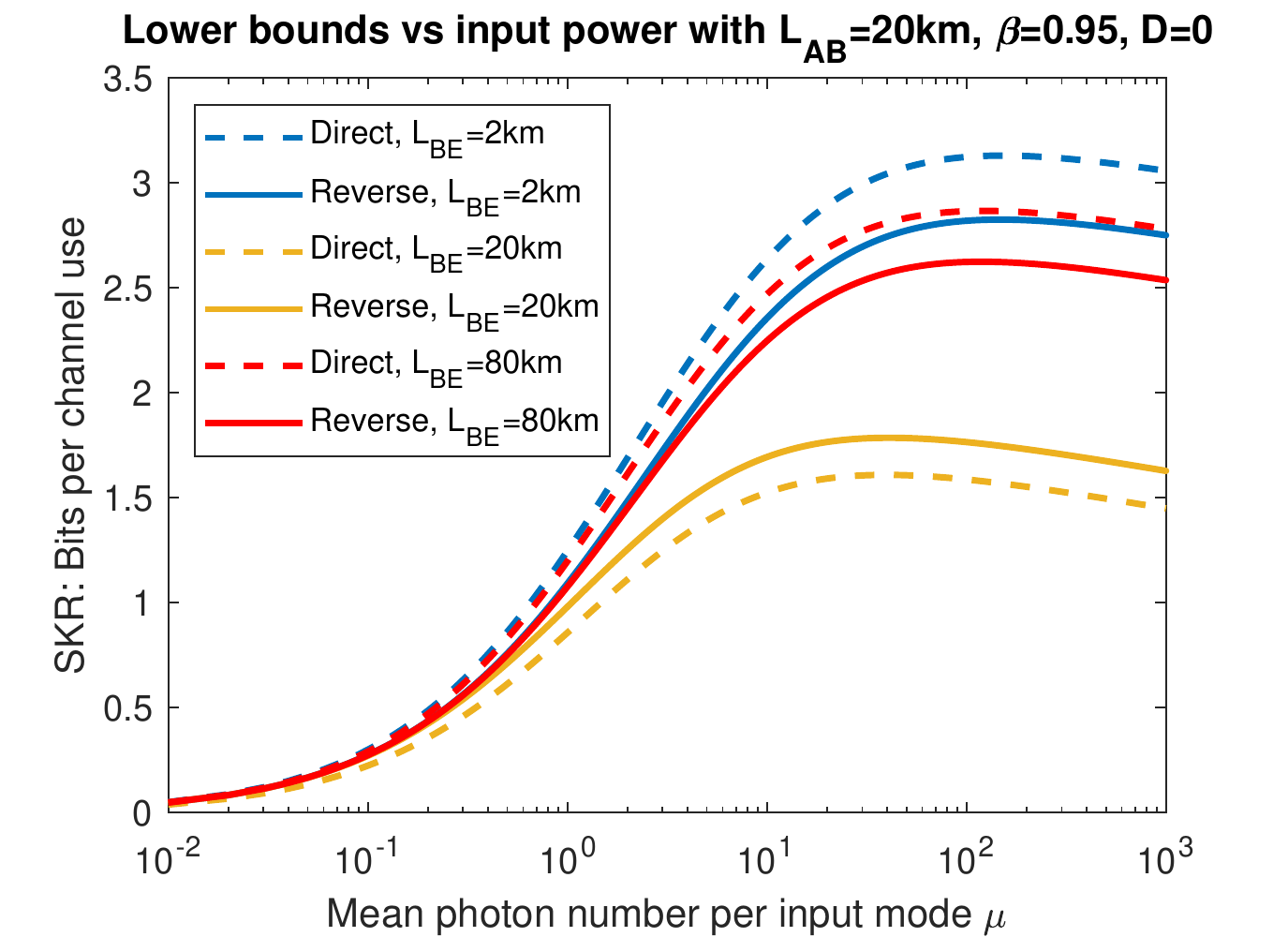}
\caption{Lower bounds vs. input power. 
The Alice-to-Bob distance $L_{AB}$ is 20km. Bob-to-Eve distances $L_{BE}$ and reconciliation schemes (direct versus reverse) are specified in the legend. Reconciliation efficiency $\beta$ is set to 0.95. Transmission center wavelength $\lambda$ is set to 1550nm. Transmitted Gaussian beam waist radius is set to $W_0=r_a=10$cm. Bob and Eve aperture radius are set to $r_b=r_e=10$cm.  \label{Index202009151945_beta95}}%ploted by Index202009151945
\end{figure}

In Fig.~\ref{Index202009151945} we plot the lower bounds versus input power with different Bob-to-Eve distances when reconciliation is perfect ($\beta=1$) and Alice-to-Bob distance $L_{AB}=20$km. For both reconciliation schemes the achievable rate lower bounds increase with increasing input power, which pushes the optimal input power to infinity. We can also see that when $L_{BE}$ increases, the achievable rate lower bounds first decrease then increase. This is due to the fact that when Eve's aperture is  aligned with Alice and Bob, her received power would first increase as the cropped Gaussian beam reconverge to its center behind Bob's aperture, then decrease as transmission loss is becoming more and more significant. When Eve's received power increases ($L_{BE}=20$km), which is equal to increasing $\kappa$, the direct reconciliation rate decreases more than reverse reconciliation although it is higher than reverse reconciliation rate either when Bob-to-Eve distance is too small ($L_{BE}=2$km) or too large ($L_{BE}=80$km).  We can see that direct reconciliation achievable rate tends to be higher than reverse reconciliation when $\kappa$ is small, similar to what we saw in~\cite{pan2019secret}. %We can also see that when Eve's aperture is small ($r_e=2\text{cm}<r_b$), the direct reconciliation (blue dashed curve) can exceed the reverse reconciliation (blue solid curve). However when Eve's aperture increases ($r_e=7\text{cm}>r_b$), the direct reconciliation (yellow dashed curve) rate drops to zero whereas reverse reconciliation (yellow solid curve) still provides a non-zero rate. This is because increasing Eve's aperture only increases Alice-to-Eve transmissivity by increasing $\kappa$ without affecting Alice-to-Bob transmissivity $\eta$, which would decrease direct reconciliation rate more, similar to what we saw in~\cite{pan2019secret}.
Next in Fig.~\ref{Index202009151945_beta95} we set the reconciliation efficiency $\beta$ to 0.95. Compared to Fig.~\ref{Index202009151945} we can see that the optimal input power for Alice is finite when reconciliation is imperfect. %However when Eve's aperture is small ($r_e=2\text{cm}<r_b$) we can achieve higher key rate exceeding this optimal input power. Similar to the perfect reconciliation case, here the direct reconciliation SKR lower bound (dashed curve) decreases to zero as Eve's aperture size increases.

%We can see from Fig.~\ref{geo1_3} that when the largest aperture accessible to both Bob and Eve increases the SKR achievable increases significantly and becomes much less sensitive to attenuation due to long transmission distance.

\subsubsection{Lower Bounds with Optimized Input Power}\label{LowerBoundswithOptimizedInputPower}

\noindent In this section we  study the lower and upper bounds with optimized input power. We first study the case with perfect reconciliation, %in Sec.~\ref{PerfectReconciliationScheme} 
taking input power  to infinity. Then with imperfect reconciliation %in Sec.~\ref{ImperfectReconciliationScheme} 
we optimize the input power numerically.

%\subsection{Perfect Reconciliation Scheme}\label{PerfectReconciliationScheme}

\begin{figure}[htbp]
\centering
\includegraphics[width=8.8cm]{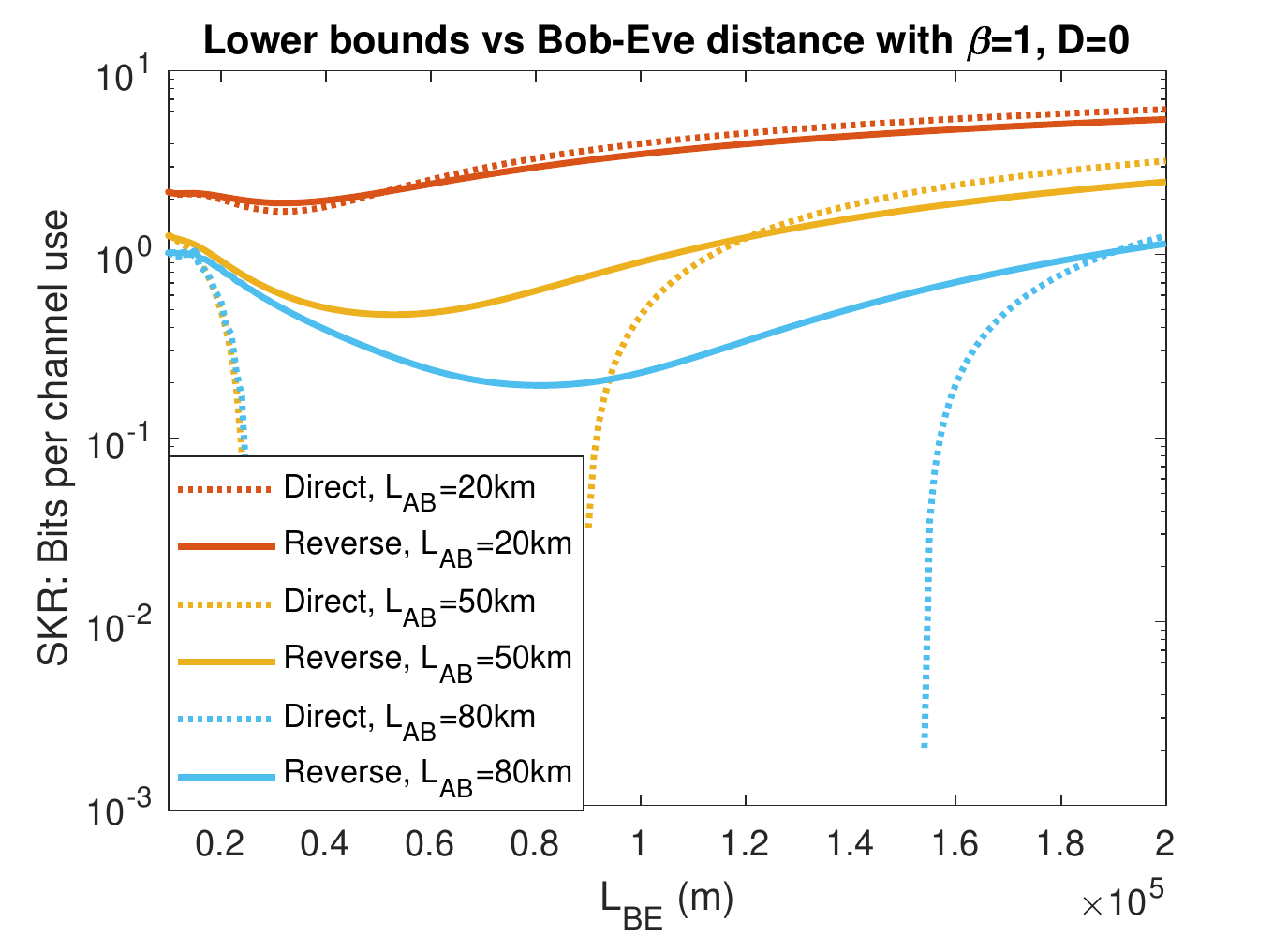}
\caption{Achievable rate direct and reverse lower bounds vs. Bob-to-Eve distance $L_{BE}$ with optimized input power (infinity as reconciliation efficiency $\beta$ is set to 1). 
 Different Alice-to-Bob distances  ($L_{AB}$) are specified in the legend.  Transmission center wavelength $\lambda$ is set to 1550nm. Transmitted Gaussian beam waist radius is set to be equal to Alice aperture radius $W_0=r_a=10$cm. Bob and Eve aperture radius are also set to $r_b=r_e=10$cm. %Here the wavelength is set to 1550nm. %the radius of Alice and Bob apertures are set to $5$\text{cm}. 
 \label{Index202007281001}}
%\end{minipage}
\end{figure}

In  Fig.~\ref{Index202007281001} we plot the direct and reverse reconciliation achievable rate lower bounds with $\beta=1$ and $\mu\rightarrow\infty$ against $L_{BE}$ assuming equal aperture sizes for Eve, Bob and Alice ($r_e=r_b=r_a=W_0=10\text{cm}$). %Here we assume the center wavelength of 1550nm. %We also include upper bounds from~\cite{pan2019secret} with dotted curves. 

%ploted by Index202007281001 in Onenote
%location:ResearchInQKD/geometric calculation/Dynamic positioning (after Bob with fixed Eve) Working code modified to fixed Eve position
%ECplotting can use Index202011111623 with data from Index202011111518

\begin{figure}[htbp]
\centering
\includegraphics[width=13cm]{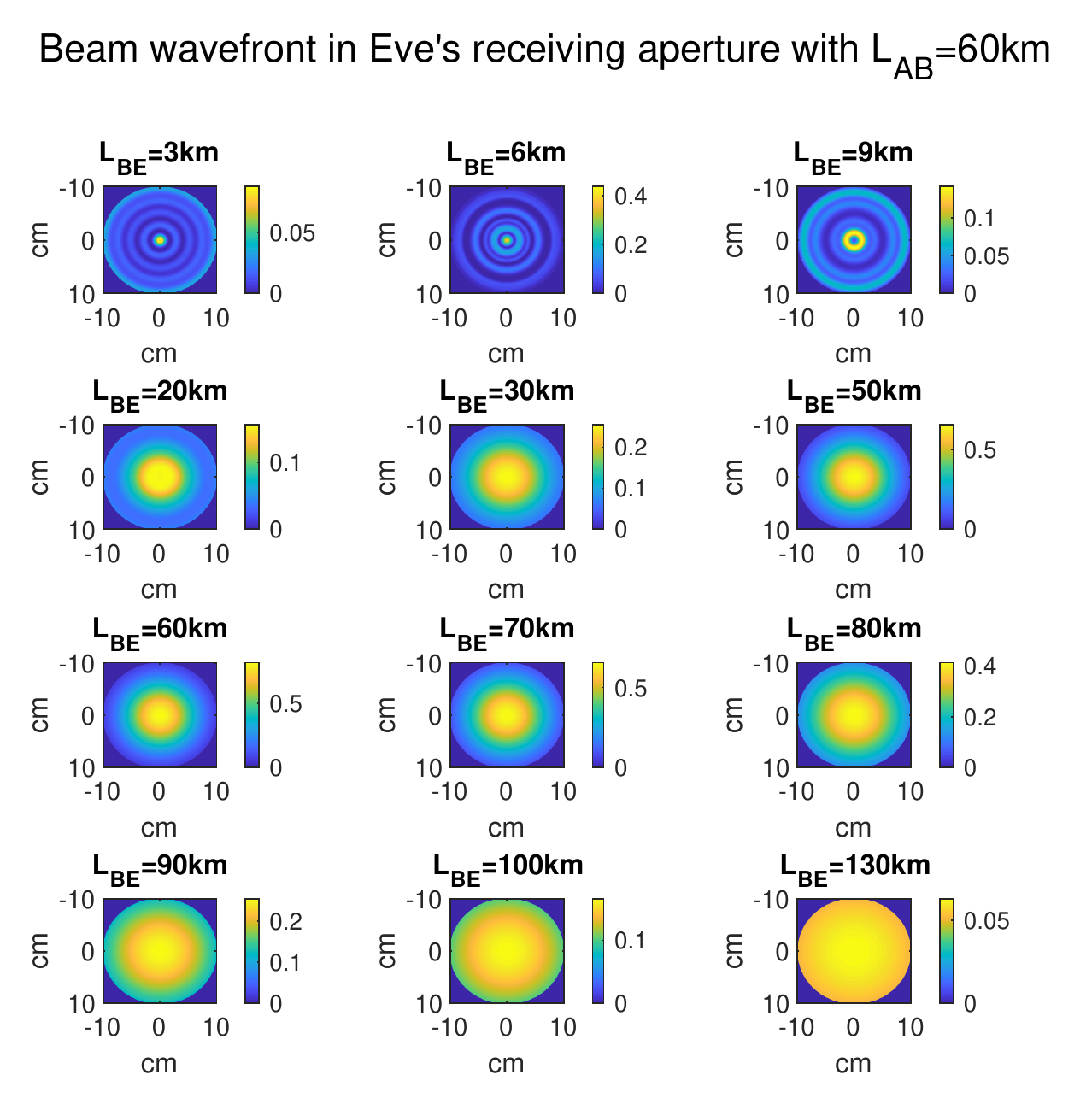}
\caption{Beam wavefront on Eve's receiving aperture with different $L_{BE}$ and $L_{AB}=60$km.
 \label{Index202011081659}}
%\end{minipage}
\end{figure}
%ploted by Index202011081659 in Onenote with data obtained with Index202010271113
%data location: OCSL-05G: C:\Users\ziwenpan\Desktop\From_old_drive_desktop\Jgariano-beam_propagator-08b531d9ee1b\Jgariano-beam_propagator-08b531d9ee1b

\begin{figure}[htbp]
\centering
\includegraphics[width=8.8cm]{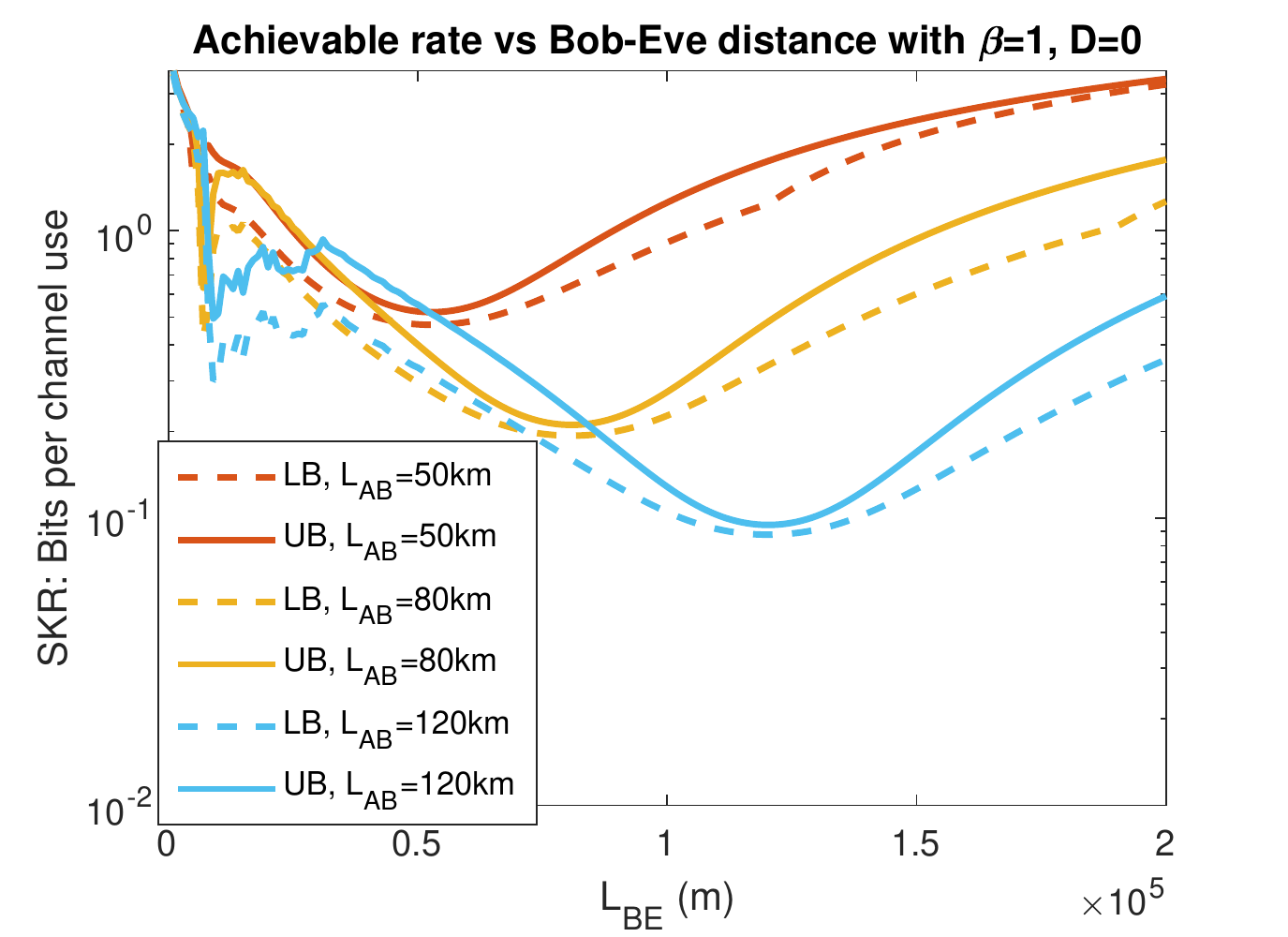}
\caption{Achievable rate lower bounds (LB) and upper bounds (UB) vs. Bob-to-Eve distance $L_{BE}$ with optimized input power (infinity as reconciliation efficiency $\beta$ is set to 1). 
 Different Alice-to-Bob distance  ($L_{AB}$) are specified in the legend.  Transmission center wavelength $\lambda$ is set to 1550nm. Transmitted Gaussian beam waist radius is set to be equal to Alice aperture radius $W_0=r_a=10$cm. Bob and Eve aperture radius are also set to $r_b=r_e=10$cm. %Here the wavelength is set to 1550nm. %the radius of Alice and Bob apertures are set to $5$\text{cm}. 
 \label{Index202008170747}}
%\end{minipage}
\end{figure}
%ploted by Index202008170747 in Onenote
%location:ResearchInQKD/geometric calculation/Dynamic positioning (after Bob with fixed Eve) Working code modified to fixed Eve position
%ECplotting can use Index202011112025 with data from Index202011111518
%attempted data set for 120km with higher precision: Index202011111518regi_0ABdistance_120km_BEdistancestep_1km_BEdistancerange_200km_rBob_10cm_w0_10cm_rEve_10cm100000registep_1.mat
%didn't get any change with that valley near 7km

In Fig.~\ref{Index202007281001} we can see that both direct and reverse reconciliation lower bounds first decrease then increase as Bob-to-Eve distance $L_{BE}$ increases. This is because the beam after Bob's aperture is a cropped Gaussian beam with a void at its center, which would gradually reconverge to its center as the propagation distance $L_{BE}$ increases, as shown in Fig.~\ref{Index202011081659}. In Fig.~\ref{Index202011081659} we take $L_{AB}=60$km as an example and showcase how the reconverging process looks like. We can see that when $L_{BE}$ is small the beam reconverging on Eve's receiving aperture is only significant within a very small region at the center while the other region is covered with constructive interference rings. % caused by constructive and destructive interference that happens during the  process. 
When $L_{BE}$ becomes larger, the beam reconverged shape starts to stablize for a range of distance and refocus to the most extent around $L_{BE}=60$km. Beyond that, the beam gradually diverges due to a large transmission distance $L_{BE}$. Since here Eve's aperture is  aligned with Alice's and Bob's, the amount of power that she can collect first increases because of the reconverging process, and then decreases as transmission loss starts to be more and more important, which caused the achievable rate in Fig.~\ref{Index202007281001} to first decrease then increase as $L_{BE}$ increases. In Fig.~\ref{Index202007281001} we can also see that as Alice-to-Bob distance $L_{AB}$ increases, the global minimum of achievable rate with Eve dynamic positioning decreases, suggesting that Eve would be able to gain more advantages by optimizing her location when $L_{AB}$ is large. We will explain this a little later. %which leads to more transmission loss to the beam as it reaches Eve, both reconciliation lower bounds decrease with the direct reconciliation rate even drops to zero at certain range of $L_{BE}$, showing that the direct reconciliation, despite its potential to achieve a higher achievable rate, is more sensitive to increased transmission loss.

Next in Fig.~\ref{Index202008170747} we plot the lower bounds as the maximum of Eqs.~(\ref{LBDmu}) and (\ref{LBRmu}) with $\beta=1$ and $\mu\rightarrow\infty$ and compare them with upper bounds from~\cite{pan2019secret} as functions of Bob-to-Eve distance $L_{BE}$ with $L_{AB}$ varied in the legend. It is worth noticing that when Eve is close to Bob there are some oscillations in the curves where the achievable rate could arrive at a local minimum, which is due to the constructive and destructive interference that happens during the process of the cropped  beam reconverging. Although the global minimum of the achievable rate decreases with increasing $L_{AB}$ like we saw in Fig.~\ref{Index202007281001}, when $L_{BE}$ is small, as $L_{AB}$ increases the achievable rate could increase  since the additional transmission loss induced by a larger $L_{AB}$ would also affect Eve especially when her location is not optimal, eg, when $L_{AB}=120$km, the achievable rate is a little higher than the $L_{AB}=80$km case when $L_{BE}$ is small. Beside that, we can also see that for each given $L_{AB}$ there is an optimal eavesdropping distance $L_{BE}$ for Eve at the  global minimum of the achievable rate where she can balance the reconverging of the cropped Gaussian beam and transmission loss to suppress the achievable  key rate between Alice and Bob. We can also see that this "optimal eavesdropping distance" increases with increasing Alice-to-Bob distance since a larger $L_{AB}$ would lead to more transmission loss and a larger beam width at Bob's aperture plane, thus taking a larger $L_{BE}$ for the beam to reconverge. Moreover, this "optimal eavesdropping distance" $L_{BE}^{\text{optimal}}$ is approximately equal to $L_{AB}$ when  $L_{AB}$ is large. To prove this, we take the Fresnel approximation and rewrite Eqs.~(\ref{PEve}) and (\ref{UEve}) as in Eqs.~(\ref{PEveFresnel}) - (\ref{UEveFresnelBessel}) below. Here in Eq.~(\ref{PEveFresnel}) we take $\phi=0$ inside the integral since when $D=0$, Eve's received light power distribution should be cylindrical symmetric. In Eq.~(\ref{UEveFresnel}) we use the Fresnel approximation and also approximate the integration region for $r$, $(r_b,\infty)$, with $\left(r_b,3W\left(L_{AB}\right)\right)$ since $\|U\left(r,L_{AB}\right)\|$ follows a Gaussian distribution.
\begin{align}
P_{Eve}&=\int_{0}^{2\pi} \int_{0}^{r_e} \|U_{Eve}\left(l,\phi\right)\|^2 \, ldld\phi=2\pi\int_{0}^{r_e} \|U_{Eve}\left(l,0\right)\|^2 \, ldl,\label{PEveFresnel}\\
    U_{Eve}\left(l,0\right)&\approx\frac{e^{ikL_{BE}}}{i\lambda L_{BE}}e^{\frac{ikl^2}{2L_{BE}}}\int_{0}^{2\pi} \int_{r_b}^{3W\left(L_{AB}\right)} U\left(r,L_{AB}\right)e^{\frac{ik}{2 L_{BE}}(r^2-2lr\cos\theta)} \, rdrd\theta,\label{UEveFresnel}\\
    &=2\pi\frac{e^{ikL_{BE}}}{i\lambda L_{BE}}e^{\frac{ikl^2}{2L_{BE}}} \int_{r_b}^{3W\left(L_{AB}\right)} U\left(r,L_{AB}\right)e^{\frac{ikr^2}{2 L_{BE}}}J_0\left(\frac{lrk}{L_{BE}}\right) \, rdr,\label{UEveFresnelBessel}
    \end{align}
Here $J_0$ denotes the zeroth order Bessel function of the first kind. The Fresnel approximation condition in our notations can be written as in Eq.~(\ref{FresnelCondition}),
\begin{align}
  L_{BE}^3&>>\left(\frac{\pi}{4\lambda}\left(l^2+r^2-2rl\cos\theta\right)^2\right)_{\text{max}},\label{FresnelCondition}\\
  &=\frac{\pi}{4\lambda}\left(l^2+9W^2\left(L_{AB}\right)+6lW(L_{AB})\right)^2,\\
  &\approx\frac{81\pi}{4\lambda}W^4(L_{AB})\label{lapprox}.
\end{align}
The approximation in Eq.~(\ref{lapprox}) is because $l$ is limited due to the size of Eve's aperture, which is small compared to $L_{BE}$. Since we have seen (and will prove) that the optimal $L_{BE}$ for Eve is approximately equal to $L_{AB}$, which is the scenario we are interested in here, %the only possible condition for violating the above restriction is with a large $W(L_{AB})$, which 
to violate the approximation condition in Eq.~(\ref{lapprox}) one would require a large $L_{AB}$ and/or a large wavelength $\lambda$ and/or a small $W_0$, eg, with $W_0=10$cm and $\lambda=1550$nm, $L_{AB}$ needs to be $4\times10^{10}$km to violate the above condition, which would not be realistic for most occasions. %long-distance transmission conditions.

Since the optimal eavesdropping distance $L_{BE}^\text{optimal}$ is near the point where the cropped Gaussian beam has re-concentrated, Eve's received power would be mostly inside a central area at  her aperture, meaning we are only concerned with small values of $l$. When $l$ is small, especially when $\frac{\sup{(lr)}k}{L_{BE}}<0.6$, we have $J_0\left(\frac{lrk}{L_{BE}}\right)\approx1$:
\begin{align}
    U_{Eve}\left(l,0\right)_{l \text{ is small}}&\approx2\pi\frac{e^{ikL_{BE}}}{i\lambda L_{BE}}e^{\frac{ikl^2}{2L_{BE}}} \int_{r_b}^{3W\left(L_{AB}\right)} U\left(r,L_{AB}\right)e^{\frac{ikr^2}{2 L_{BE}}} \, rdr,\label{J0_1}\\
    &=M\int_{r_b}^{3W\left(L_{AB}\right)}e^{\frac{-r^2}{W^2\left(L_{AB}\right)}-\frac{ikr^2}{2R(L_{AB})}+\frac{ikr^2}{2 L_{BE}}} \, rdr,\\
    M&=2\pi\frac{e^{ikL_{BE}}}{i\lambda L_{BE}}e^{\frac{ikl^2}{2L_{BE}}}\frac{E_0W_0}{W(L_{AB})}e^{-i\left(kL_{AB}-\psi\left(L_{AB}\right)\right)}.
\end{align}
Thus we have:
\begin{align}
    \left\|U_{Eve}\left(l,0\right)_{l \text{ is small}}\right\|&\approx\frac{2\pi E_0 W_0}{\lambda L_{BE}W(L_{AB})}\left\|\int_{r_b}^{3W\left(L_{AB}\right)}e^{\frac{-r^2}{W^2\left(L_{AB}\right)}-\frac{ikr^2}{2R\left(L_{AB}\right)}+\frac{ikr^2}{2 L_{BE}}} \, rdr\right\|\\
    &=\frac{2E_0W_0}{W(L_{AB})}\left\|\frac{k}{2L_{BE}}\int_{r_b}^{3W\left(L_{AB}\right)}e^{\frac{-1}{W^2\left(L_{AB}\right)}r^2+i\left(\frac{k}{2 L_{BE}}-\frac{k}{2R\left(L_{AB}\right)}\right)r^2} \, rdr\right\|\\
    &=\frac{E_0W_0}{W(L_{AB})}\left\|\frac{k}{2L_{BE}}\int_{r_b}^{3W\left(L_{AB}\right)}e^{\left(\frac{-1}{W^2\left(L_{AB}\right)}+i\left(\frac{k}{2 L_{BE}}-\frac{k}{2R\left(L_{AB}\right)}\right)\right)r^2} \, 2rdr\right\|\\
    &=\frac{E_0W_0}{W(L_{AB})}\left\|B\int_{r_b^2}^{D}e^{\left(A+i\left(B-C\right)\right)x} \, dx\right\|\\
    &=\frac{E_0W_0}{W(L_{AB})}\left\|\frac{B}{A+i(B-C)}\left(e^{\left(A+i\left(B-C\right)\right)D}-e^{\left(A+i\left(B-C\right)\right)r_b^2}\right)\right\|\\
    &=\frac{E_0W_0}{W(L_{AB})}\frac{1}{\sqrt{\left(\frac{A}{B}\right)^2+\left(1-\frac{C}{B}\right)^2}}\left\|\left(e^{AD}e^{i\left(B-C\right)D}-e^{Ar_b^2}e^{i\left(B-C\right)r_b^2}\right)\right\|\\
    &=\frac{E_0W_0}{W(L_{AB})}f_1(B)f_2(B)
    \end{align}
    with
    \begin{align}
    A&=\frac{-1}{W^2\left(L_{AB}\right)}\\
    B&=\frac{k}{2L_{BE}}\\
    C&=\frac{k}{2R\left(L_{AB}\right)}\\
    D&=9W^2\left(L_{AB}\right)=9W_0^2\left(1+\left(\frac{L_{AB}}{z_0}\right)^2\right)=9W_0^2\left(1+\left(\frac{\lambda L_{AB}}{W_0^2\pi}\right)^2\right)\\
    f_1(B)&=\frac{1}{\sqrt{\left(\frac{A}{B}\right)^2+\left(1-\frac{C}{B}\right)^2}}=\frac{1}{\sqrt{\frac{\lambda ^2 (L_{BE}-L_{AB})^2+\pi ^2 W_0^4}{\lambda ^2 L_{AB}^2+\pi ^2 W_0^4}}}\label{f1}\\
    f_2(B)&=\left\|\left(e^{AD}e^{i\left(B-C\right)D}-e^{Ar_b^2}e^{i\left(B-C\right)r_b^2}\right)\right\|\label{f2}
\end{align}

To optimize $f_1(B)$ in Eq.~(\ref{f1}), we have 
\begin{align}
    \arg\max_{L_{BE}} f_1(B)=L_{AB} 
\end{align}

To optimize $f_2(B)$ in Eq.~(\ref{f2}), we have:
\begin{align}
    \left(\arg\max_{B} f_2(B)-C\right)D-\left(\arg\max_{B} f_2(B)-C\right)r_b^2&=(2n+1)\pi \text{  }\text{  }\text{  }\text{  }\text{  }\text{  }\text{  }\text{  }\text{  }\text{  }\text{  }\text{  }\text{  }\text{  }\text{  }(n\in\mathbb{Z})\\
    \arg\max_{B} f_2(B)&=C+\frac{(2n+1)\pi}{D-r_b^2}\text{  }\text{  }\text{  }\text{  }\text{  }\text{  }\text{  }(n\in\mathbb{Z})
\end{align}
which gives us:
\begin{align}
    \arg\max_{L_{BE}} f_2(B)=\frac{1}{\lambda  \left(\frac{2 n+1}{9 \left(\frac{\lambda ^2 L_{AB}^2}{\pi ^2 W_0^2}+W_0^2\right)-r_b^2}+\frac{\lambda  L_{AB}}{\lambda ^2 L_{AB}^2+\pi ^2 W_0^4}\right)}\text{  }\text{  }\text{  }\text{  }\text{  }(n\in\mathbb{Z})
\end{align}
If we set $n=0$, when $L_{AB}$ increases, the first term inside the bracket in the denominator $\frac{2 n+1}{9 \left(\frac{\lambda ^2 L_{AB}^2}{\pi ^2 W_0^2}+W_0^2\right)-r_b^2}$ decreases faster than the second term $\frac{\lambda  L_{AB}}{\lambda ^2 L_{AB}^2+\pi ^2 W_0^4}$ and eventually goes to zero when $L_{AB}$ is large. As a result, when $n=0$, we have:
\begin{align}
    \arg\max_{L_{BE}} f_2(B)_{L_{AB} \text{ is large}}\approx \frac{\lambda ^2 L_{AB}^2+\pi ^2 W_0^4}{\lambda^2 L_{AB}}\approx L_{AB}=\arg\max_{L_{BE}} f_1(B)
\end{align}
To conclude, we have:
\begin{align}
     \arg\max_{L_{BE}} \| U_{Eve}\left(l,0\right)\|_{l \text{ is small, }L_{AB} \text{ is large}}\approx L_{AB} \label{OptimalEavesdroppingDistance}
\end{align}
which means that for a large Alice-to-Bob distance and relatively small apertures, the optimal eavesdropping distance for Eve $L_{BE}^\text{optimal}\approx L_{AB}$, which is consistent with what we observed. 

\begin{figure}[htbp]
\centering
\includegraphics[width=8.8cm]{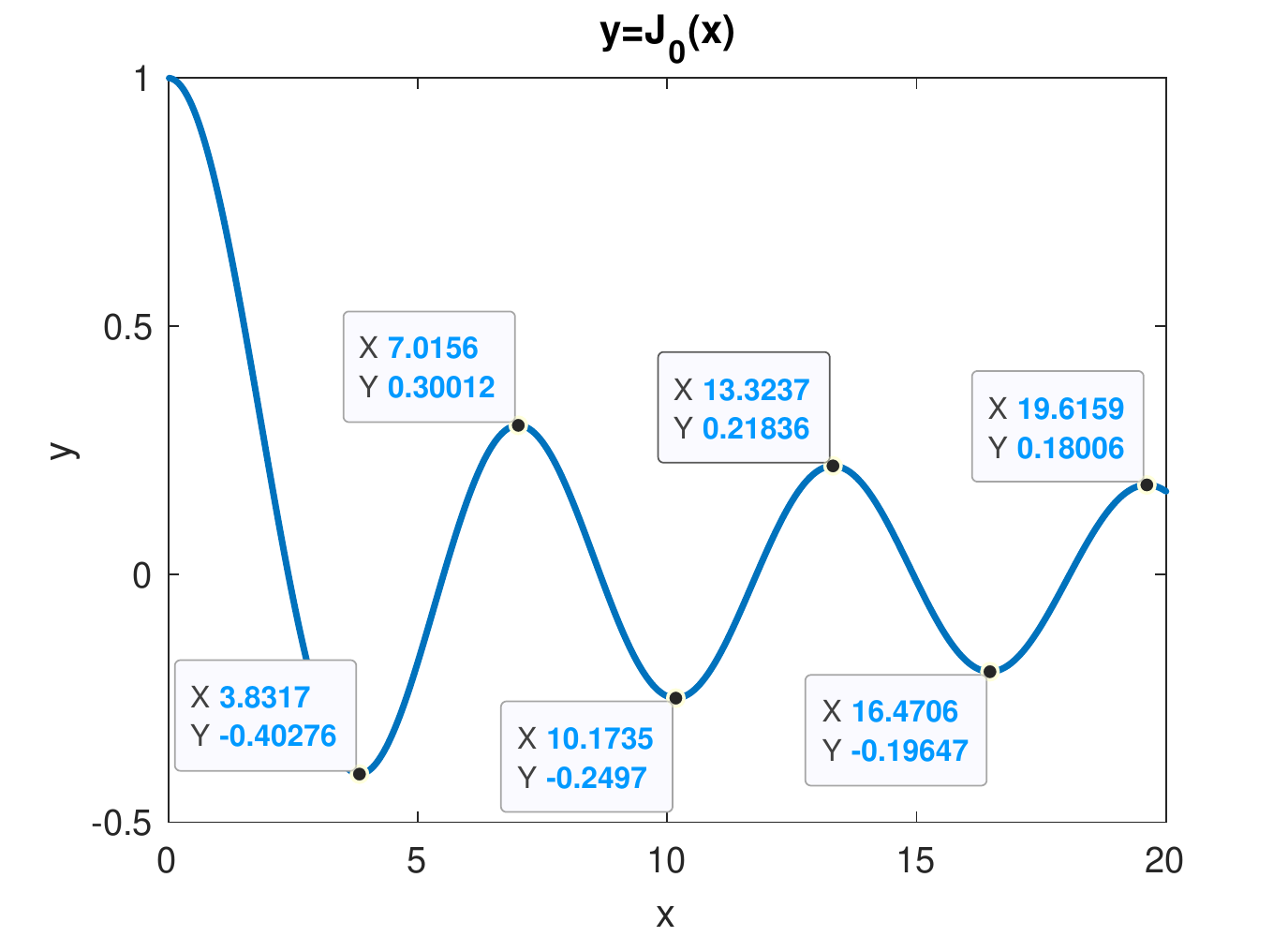}
\caption{Zeroth order Bessel function of the first kind $J_0(x)$. %Here the wavelength is set to 1550nm. %the radius of Alice and Bob apertures are set to $5$\text{cm}. 
 \label{Index202012181601}}
%\end{minipage}
\end{figure}

When $L_{BE}$ is small, the approximation condition  $J_0\left(\frac{lrk}{L_{BE}}\right)\approx1$ with $\frac{\sup{(lr)}k}{L_{BE}}<0.6$ used in Eq.~(\ref{J0_1}) would not be valid. If the condition is relaxed to a lesser degree, eg, $\frac{\sup{(lr)}k}{L_{BE}}<1$, the conclusion we obtained should still work but with a larger deviation. If $L_{BE}$ further decreases, Eq.~(\ref{J0_1}) and what follows would not be valid and some local minimums of the achievable rate, like in Fig.~\ref{Index202008170747}, could emerge when the integration of Eq.~(\ref{UEveFresnelBessel}) is around the peak of $\left\|J_0\left(\frac{lrk}{L_{BE}}\right)\right\|$ as marked in Fig.~\ref{Index202012181601}. For example, when $L_{AB}$ is not too large so that the approximated integration region $\left(r_b, 3W\left(L_{AB}\right)\right)$ is not too large, since the cropped Gaussian beam would have the most power near its cropped edge ($r=r_b$), which would occupy a larger portion in the integration over $\left(r_b, 3W\left(L_{AB}\right)\right)$ than the region away from the edge, the local minimums of the achievable rate could emerge near $L_{BE}=\frac{r_b r_e k}{3.8317}$ and/or $L_{BE}=\frac{r_b r_e k}{7.0156}$ which are approximately 10.5km and 5.8km with $\lambda=1550$nm, $r_b=r_e=10$cm, same as what we observed. When $L_{AB}$ becomes larger, since the approximated integration region $\left(r_b, 3W\left(L_{AB}\right)\right)$ is larger and can include many peaks of $\left\|J_0\left(\frac{lrk}{L_{BE}}\right)\right\|$, the local minimums of the achievable rate could shift around depending on specific values of $L_{AB}$. 

%ploted by Index202012181601 in Onenote 

Next if we look at $\left\|U_{Eve}\left(l,0\right)\right\|$ with $L_{BE}=L_{AB}$:
\begin{align}
    &\left\|U_{Eve}\left(l,0\right)\right\|_{l\text{ is small, }L_{BE}=L_{AB}}\approx\nonumber\\
    &\left\|E_0 \sqrt{\frac{W_0^4}{\lambda ^2 L_{AB}^2+\pi ^2 W_0^4}} \sqrt{\frac{\lambda ^2 L_{AB}^2}{W_0^4}+\pi ^2} e^{-\frac{\pi ^2 W_0^2 r_b^2}{\lambda ^2 L_{AB}^2+\pi ^2 W_0^4}-9} \left(e^{\frac{\pi ^2 W_0^2 r_b^2}{\lambda ^2 L_{AB}^2+\pi ^2 W_0^4}}-e^9\right)\right\|\\
    &\approx E_0(1-e^{-9})\text{  }\text{  }\text{  }\text{  }\text{  }\text{  }\text{  }\text{  }\text{  }\text{  }\text{  }\text{  }(L_{AB}\text{ is large})
\end{align}

\begin{figure}[htbp]
\centering
\includegraphics[width=8.8cm]{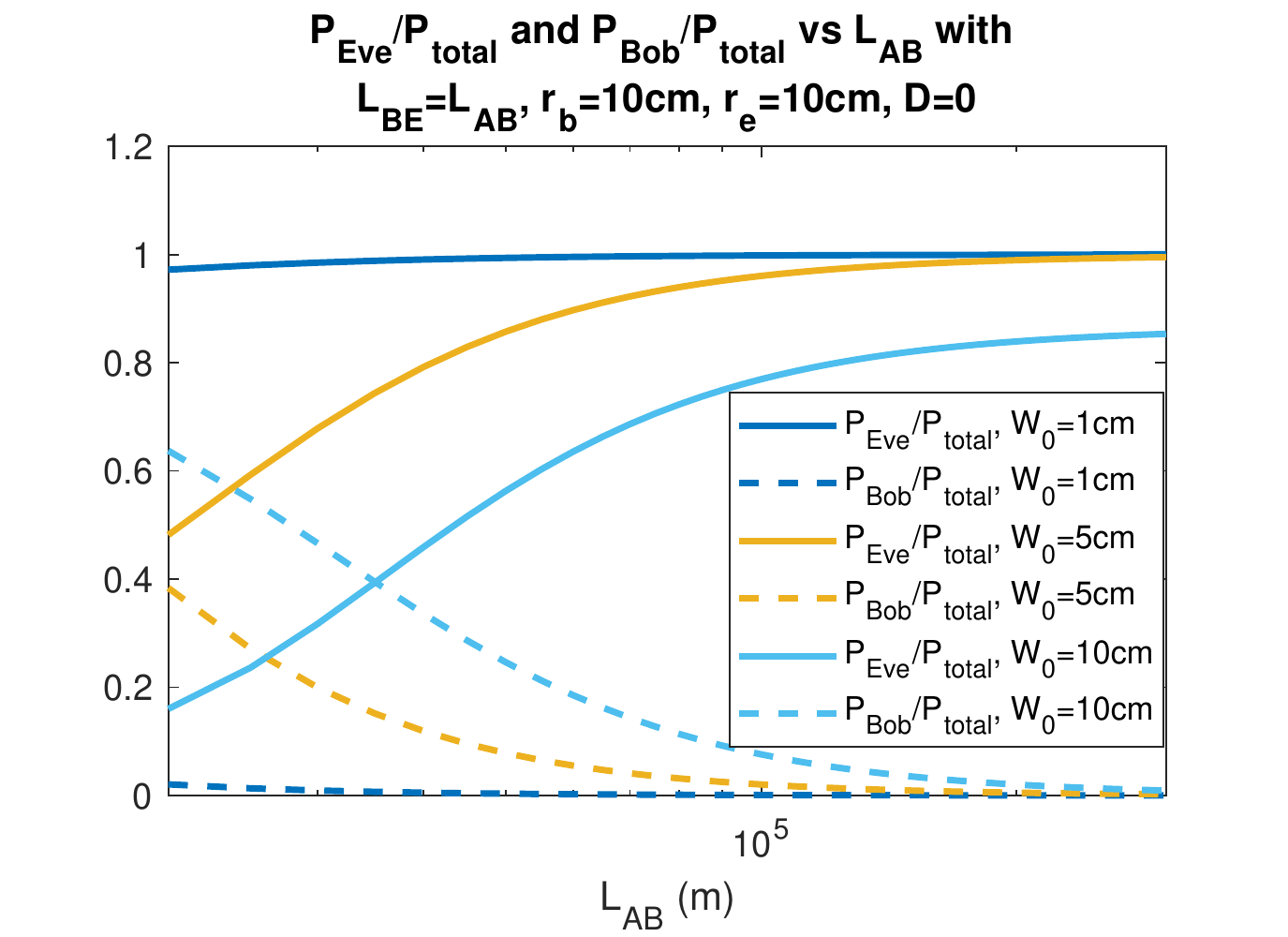}
\caption{$P_{Eve}/P_{total}$ and $P_{Bob}/P_{total}$ vs. Alice-to-Bob distance $L_{AB}$ with $L_{BE}=L_{AB}$. 
 Different $W_0$ are specified in the legend.  Transmission center wavelength $\lambda$ is set to 1550nm.  Bob and Eve aperture radius are also set to $r_b=r_e=10$cm. %Here the wavelength is set to 1550nm. %the radius of Alice and Bob apertures are set to $5$\text{cm}. 
 \label{PEvevsPAliceDoublelogscale}}
%\end{minipage}
\end{figure}
%ploted by Index202011112219 in Onenote with data from Index202011112100

In Fig.~\ref{PEvevsPAliceDoublelogscale} we plot the ratio of $\frac{P_{Eve}}{P_{total}}$ and $\frac{P_{Bob}}{P_{total}}$ as functions of $L_{AB}$ with different $W_0$ while $L_{BE}$ is set to be equal to $L_{AB}$. We can see that as $L_{AB}$ increases, $\frac{P_{Eve}}{P_{total}}$ increases since a larger $L_{AB}$ would fit more with our approximation conditions in deriving the optimal eavesdropping distance $L_{BE}=L_{AB}$, so the beam would be more reconverged at that point. And when $W_0$ is small, $\frac{P_{Eve}}{P_{total}}$  even becomes close to one as $L_{AB}$ increases, meaning Eve is receiving almost all of the power from the transmitted Gaussian beam while Bob receives almost none due to a larger divergence angle induced by smaller $W_0$. % as is shown. %in Fig.~\ref{PBobvsPAliceDoublelogscale}. 
However, it is interesting to see that in the far-field regime the Gaussian beam with a larger divergence angle is able to become more refocused near the point where $L_{BE}=L_{AB}$, which means that Eve can receive a much larger portion of the transmitted power than Bob in a long-distance transmission scenario ($L_{AB}$ is large) by moving her aperture to the optimal eavesdropping distance $L_{BE}^\text{optimal}$. This explains why the global minimum of the achievable rate with eavesdropper dynamic positioning decreases with increasing $L_{AB}$ in Fig.~\ref{Index202007281001} and Fig.~\ref{Index202008170747}.

\begin{comment}
\begin{figure}[htbp]
\centering
\includegraphics[width=8.8cm]{PBobvsPAliceDoublelogscale.pdf}
\caption{$P_{Bob}/P_{total}$ vs. Alice-to-Bob distance $L_{AB}$. 
 Different $W_0$ are specified in the legend.  Transmission center wavelength $\lambda$ is set to 1550nm.  Bob and Eve aperture radius are also set to $r_b=r_e=10$cm. %Here the wavelength is set to 1550nm. %the radius of Alice and Bob apertures are set to $5$\text{cm}. 
 \label{PBobvsPAliceDoublelogscale}}
%\end{minipage}
\end{figure}
%ploted by Index202011112219 in Onenote with data from Index202011112100
\end{comment}

\begin{figure}[htbp]
\centering
\includegraphics[width=8.8cm]{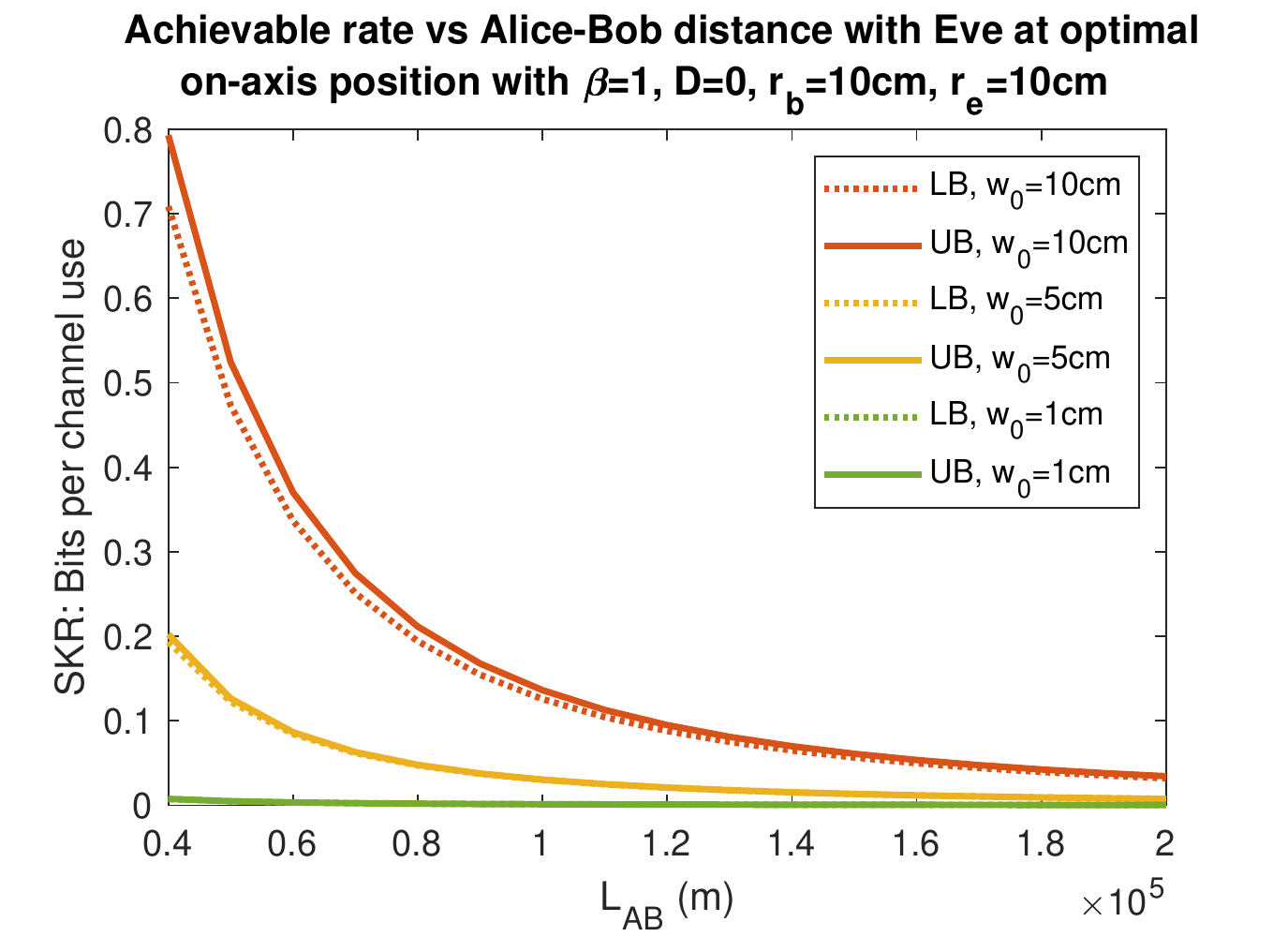}
\caption{Lower bounds (LB) and upper bounds (UB) over achievable rate vs. Alice-to-Bob distance $L_{AB}$ with Eve located at optimal on-axis position ($L_{BE}=L_{AB}$).
 Different $W_0$ are specified in the legend.  Transmission center wavelength $\lambda$ is set to 1550nm.  Bob and Eve aperture radius are also set to $r_b=r_e=10$cm. %Here the wavelength is set to 1550nm. %the radius of Alice and Bob apertures are set to $5$\text{cm}. 
 \label{Index202010142350}}
%\end{minipage}
\end{figure}
%ploted by Index202010142350 in Onenote with data from Index202010142313

Next we look at the achievable rate lower and upper bounds with $L_{BE}=L_{AB}$. In Fig.~\ref{Index202010142350} we can see that the upper bounds are really close to the lower bounds when $L_{BE}=L_{AB}$ as we saw with previous figures, especially when $L_{AB}$ is large. %since $\kappa$ would be large
The achievable rate decreases as $L_{AB}$ increases since Bob would receive a less portion of the transmitted power while Eve can collect more (corresponding to a smaller $\eta$   and a larger $\kappa$), similar to what we saw in Fig.~\ref{Index202007281001} and Fig.~\ref{Index202008170747}. As the beam waist radius $W_0$ increases the achievable rate increases because the beam divergence angle would be smaller, which would let Bob collect more power in this case.

\begin{figure}[htbp]
\centering
\includegraphics[width=8.8cm]{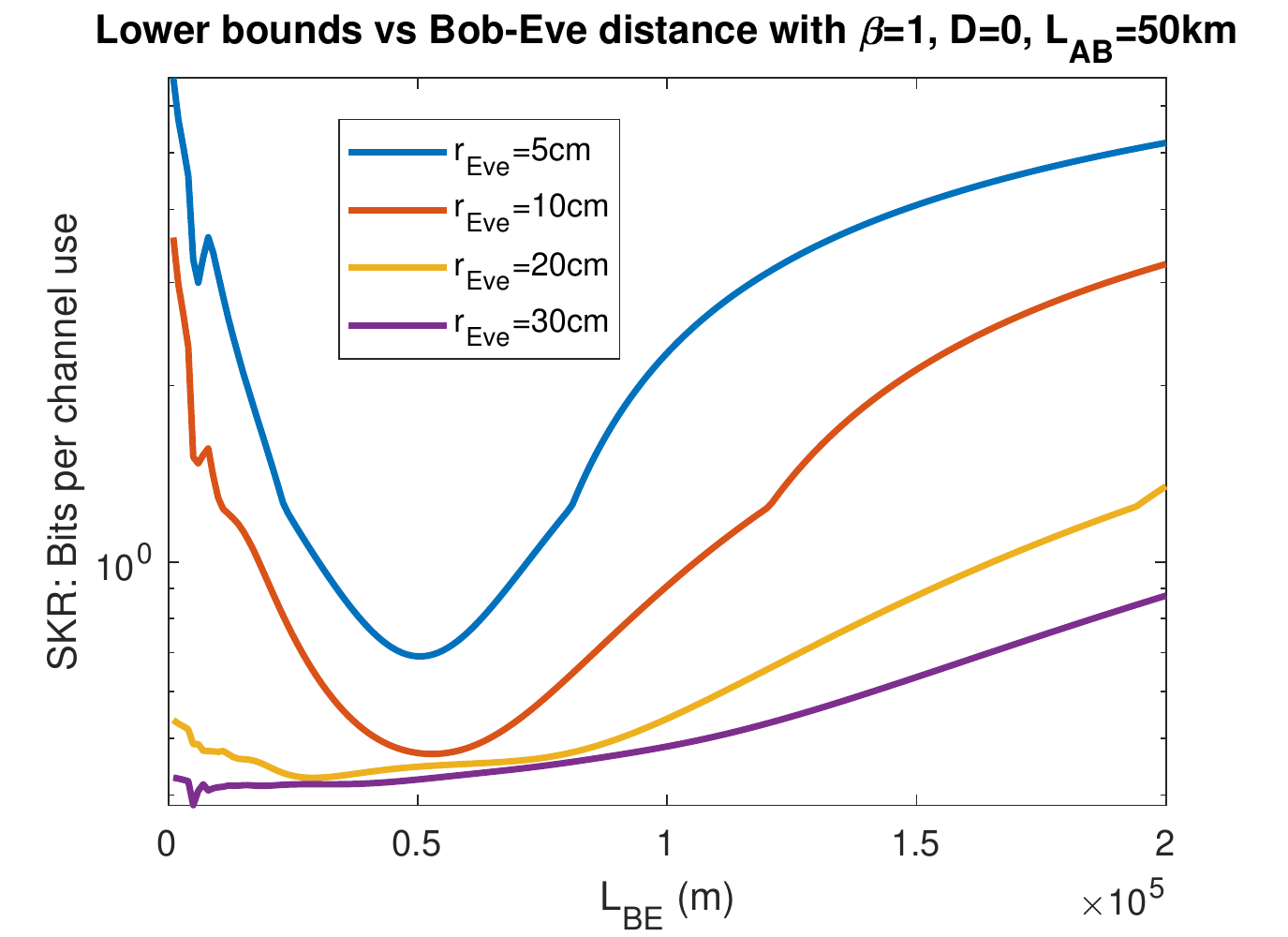}
\caption{Lower bounds versus $L_{BE}$ with $L_{AB}=50$km, $W_0=10$cm and $r_{Bob}=10$cm. Different $r_{Eve}$ are specified in the legend.  Transmission center wavelength $\lambda$ is set to 1550nm.  
 \label{AB50kmrEveVaried}}
%\end{minipage}
\end{figure}
%ploted by Index202011211417 in Onenote with data from 202011141527 

Since we assumed $l$ to be small in the above derivation for Eve's optimal eavesdropping distance, which means $r_{Eve}$ is not large, we now look at how increasing $r_{Eve}$ would affect the achievable  key rate lower bounds. As we have shown in Fig.~\ref{Index202011081659} that the optimal eavesdropping distance exists because the reconverging process of the cropped Gaussian beam to some extent counters the effect of the transmission loss, it would be reasonable to deduct that when Eve has a larger aperture, she should be able to get closer to Bob to collect more power even before the beam reconverging process is complete. In Fig.~\ref{AB50kmrEveVaried}  we fix $L_{AB}=50$km and plot the achievable rate lower bounds as the maximum of direct and reverse lower bounds with only $r_{Eve}$ is varied. We can see that when $r_{Eve}$ increases from 5cm to 10cm the optimal eavesdropping distance $L_{BE}^\text{optimal}$ doesn't change much although the achievable rate decreases as Eve is able to collect more photons with a larger aperture. When $r_{Eve}$ further increases to 20cm and 30cm we can see that the lower bound curve has been more flattened along the x-axis. And we can see that it is possible for Eve to achieve better eavesdropping effect when $L_{BE}$ is smaller than $L_{AB}$. However, we can also see that the additional advantage gained by Eve with a larger aperture is minimal since the power collecting advantage only comes from a lower channel loss when Eve approaches closer to Bob and uses a large enough aperture to fully collect the unfocused beam.

\begin{figure}[htbp]
\centering
\includegraphics[width=8.8cm]{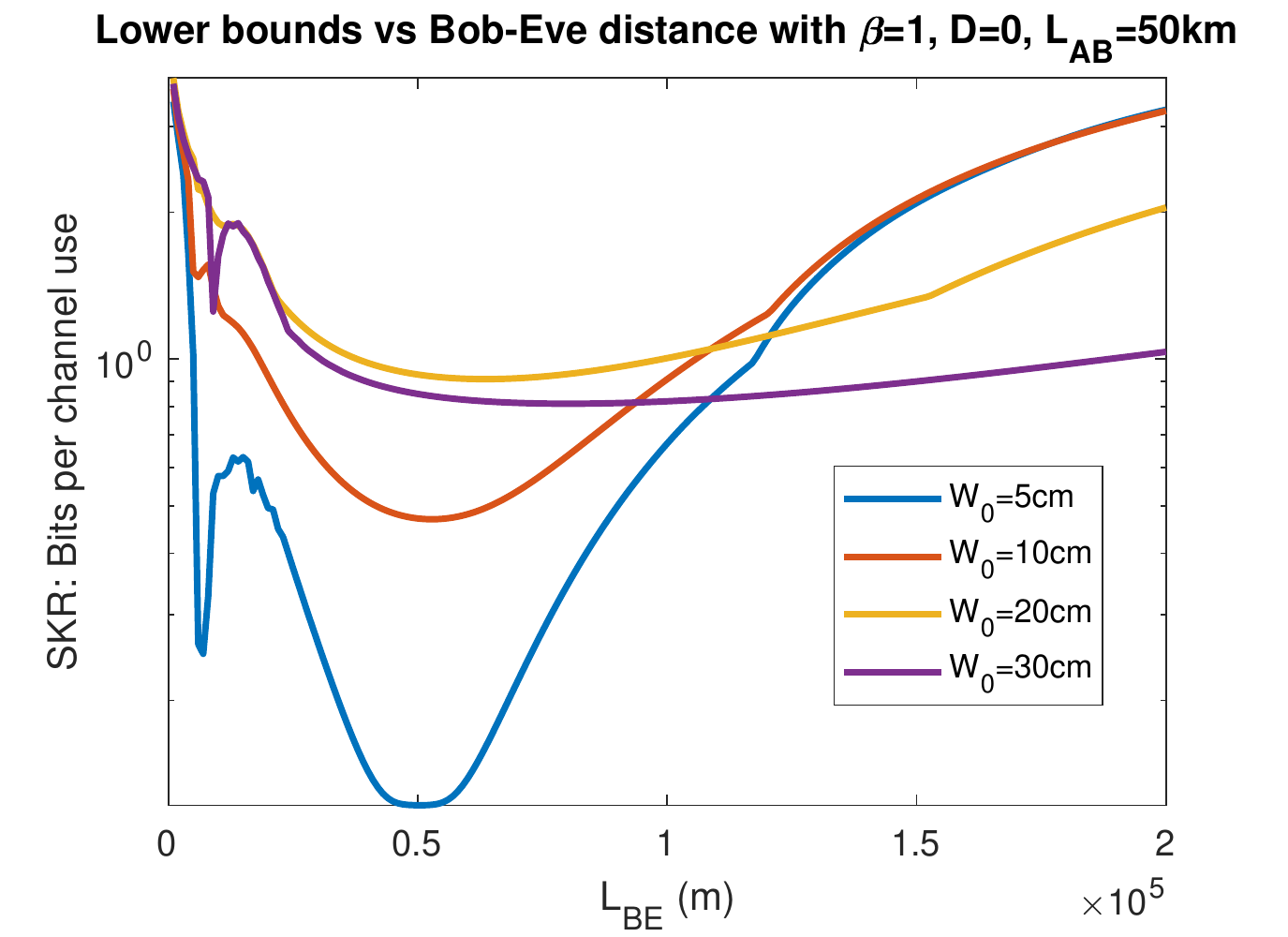}
\caption{Lower bounds versus $L_{BE}$ with $L_{AB}=50$km, $r_{Eve}=10$cm and $r_{Bob}=10$cm. Different $W_0$ are specified in the legend.  Transmission center wavelength $\lambda$ is set to 1550nm.%Here the wavelength is set to 1550nm. %the radius of Alice and Bob apertures are set to $5$\text{cm}. 
 \label{AB50kmW0Varied}}
%\end{minipage}
\end{figure}
%ploted by Index202011211449 in Onenote with data from 202011141527 

Next in Fig.~\ref{AB50kmW0Varied} we fix Eve's aperture radius as 10cm and vary $W_0$. When $W_0$ increases from 5cm to 10cm, since an increased $W_0$ decreases the beam's divergence angle, making it more focused so Bob can receive more power, we can see that the achievable rate  increases but still has a global minimum value around the point where $L_{BE}=L_{AB}$. However, when $W_0$ further increases, the achievable rate curve starts to flatten out when $L_{BE}$ is large. This is because when $W_0$ further increases, the beam Rayleigh length increases to the point that $L_{AB}$ cannot be viewed as relatively large anymore so the approximation that we used in deriving Eq.~(\ref{OptimalEavesdroppingDistance}) is no longer valid.

Now that we have good characterization for the case when $L_{AB}$ is large,  it is worth looking into the case when $L_{AB}$ is small. In that case, when the Gaussian beam Rayleigh length is large compared to the transmission distance $L_{AB}$, the beam transmitted can be viewed as approximately collimated~\cite{fischer2007dark}, which mimics the signature of a point source and thus can be better characterized with existing results from the famous Arago spot~\cite{harvey1984spot}. In~\cite{harvey1984spot,reisinger2017relative} the relative transverse complex amplitude distribution $U^{rel}$ compared to the undisturbed wavefront was pointed out for an ideal point source, which we rewrite in our notations here:
\begin{align}
    U^{rel}\left(l,L_{BE}\right)&=-\sqrt{\frac{L_{BE}^2}{L_{BE}^2+r_b^2}}J_0\left(\frac{2r_b\pi l}{\lambda L_{BE}}\right)e^{ikL_{BE}+ik\frac{l^2+r_b^2}{2L_{BE}}},\\
    \left\|U^{rel}\left(l,L_{BE}\right)\right\|&=\sqrt{\frac{L_{BE}^2}{L_{BE}^2+r_b^2}}\left\|J_0\left(\frac{2r_b\pi l}{\lambda L_{BE}}\right)\right\|.\label{RelTranAmpDis}
\end{align}
%\textbf{Does this lead to possible explanations on the local minimum before global minimum emerges? Perhaps when $J_0$ takes its minimum value? } 
Below we use this relative transverse amplitude distribution in Eq.~(\ref{RelTranAmpDis}) and the undisturbed wavefront to try to predict power reception of Bob and Eve (shown in dashed curves) and we show that this method works well specifically when the transmitted beam is approximately collimated, which is either when $L_{AB}$ is small and/or $W_0$ is large. 
%Here we use the results from~\cite{reisinger2017relative} with its Eq.~(1) rewritten here as:\textbf{There is probably an error here since I was using the relative intensity on every point of the Gaussian beam, but now it seems like that only works with the peak intensity, with different positions}
%\begin{align}
    %&\textbf{The problem here is, whether to use Arago spot to model Gaussian beam, }\\
    %&\textbf{or directly use ideal point source for comparison. }\\
    %&\textbf{See code Index202011122308 and Index202011121746 for reference.}
%\end{align}
%To have a better understanding of this, we compare our results with what can be obtained from the famous Arago spot~\cite{harvey1984spot}, which can be used to well characterize a point source for Alice. 

\begin{figure}[htbp]
\centering
\includegraphics[width=8.8cm]{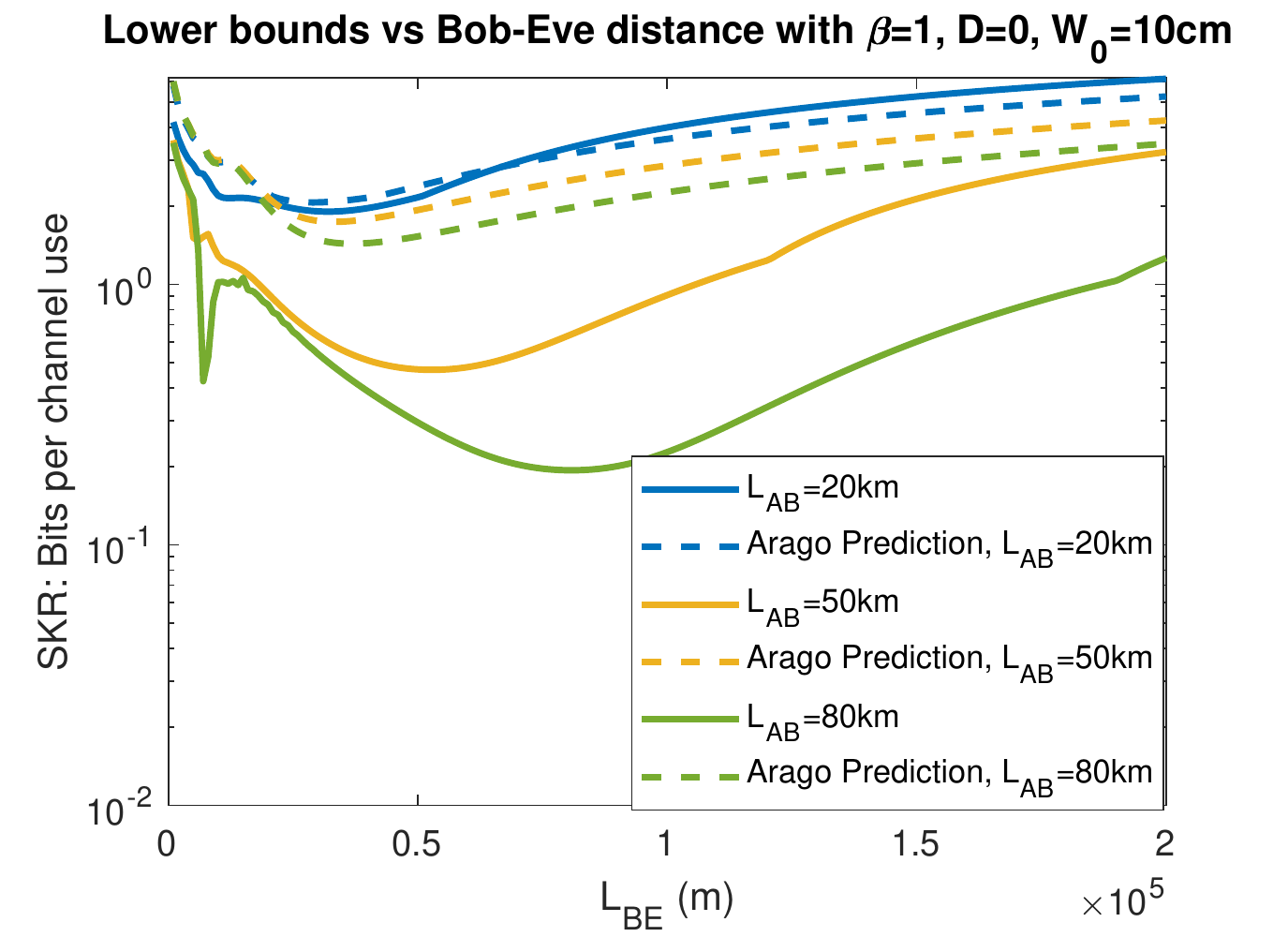}
\caption{Lower bounds versus $L_{BE}$ for different values of $L_{AB}$ compared with the predicted lower bounds using the Arago spot result in dashed curves. 
  Transmission center wavelength $\lambda$ is set to 1550nm.  Bob and Eve aperture radius are  set to $r_b=r_e=10$cm. $W_0=10$cm. %Here the wavelength is set to 1550nm. %the radius of Alice and Bob apertures are set to $5$\text{cm}. 
 \label{Index202011121746CompwithAragoSpot}}
%\end{minipage}
\end{figure}
%ploted by Index202011121746 in Onenote with data from Index202011112100

\begin{figure}[htbp]
\centering
\includegraphics[width=8.8cm]{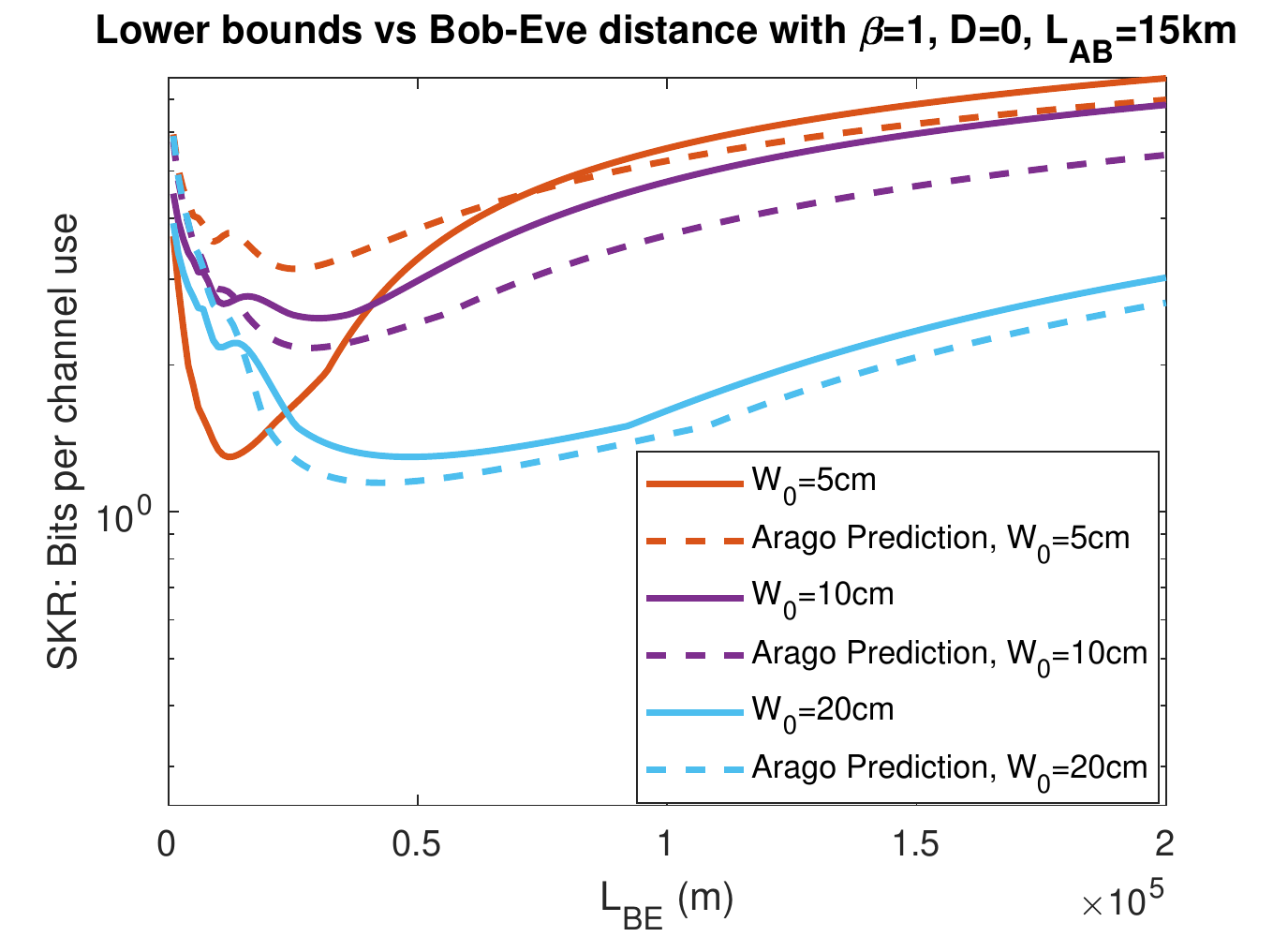}
\caption{Lower bounds versus $L_{BE}$ for different values of $W_0$ compared with the predicted lower bounds using the Arago spot result in dashed curves. 
  Transmission center wavelength $\lambda$ is set to 1550nm.  Bob and Eve aperture radius are  set to $r_b=r_e=10$cm. $L_{AB}=15$km.  %Here the wavelength is set to 1550nm. %the radius of Alice and Bob apertures are set to $5$\text{cm}. 
 \label{Index202011122308}}
%\end{minipage}
\end{figure}
%ploted by Index202011122308 in Onenote with data from Index202011112100

As  in Fig.~\ref{Index202011121746CompwithAragoSpot}, we include results on the achievable rate lower bounds versus $L_{BE}$ with different $L_{AB}$ specified in the legend and compare them with the predicted results in dashed curves using the Arago Spot theory (Arago Prediction). When $L_{AB}$ decreases the lower bound curves become closer to the Arago Prediction curves as when $L_{AB}$ is small the Gaussian beam can be viewed as approximately collimated so it mimics the signature of a point source. This is even clearer if we tune the value of $W_0$. In Fig.~\ref{Index202011122308} we fix $L_{AB}=15$km and vary the value of $W_0$. When $W_0$ is small (5cm), it's clear that the achievable rate  reaches its global minimum around the point where $L_{BE}=L_{AB}$. However, as $W_0$ increases, the achievable rate curves are closer to the Arago Prediction curves  since an increased Gaussian beam waist radius decreases beam divergence, making it more collimated so it is more similar to the point source. %\textbf{Something is wrong here, the $W_0$ is listed for Gaussian beams, meaning here the point source is still modeling the Gaussian beam with the Arago Spot result.}

\begin{figure}[htbp]
\centering  %图片全局居中
\subfigure[]{
\label{Index202012141329_1_50km_D0}
\includegraphics[width=0.45\textwidth]{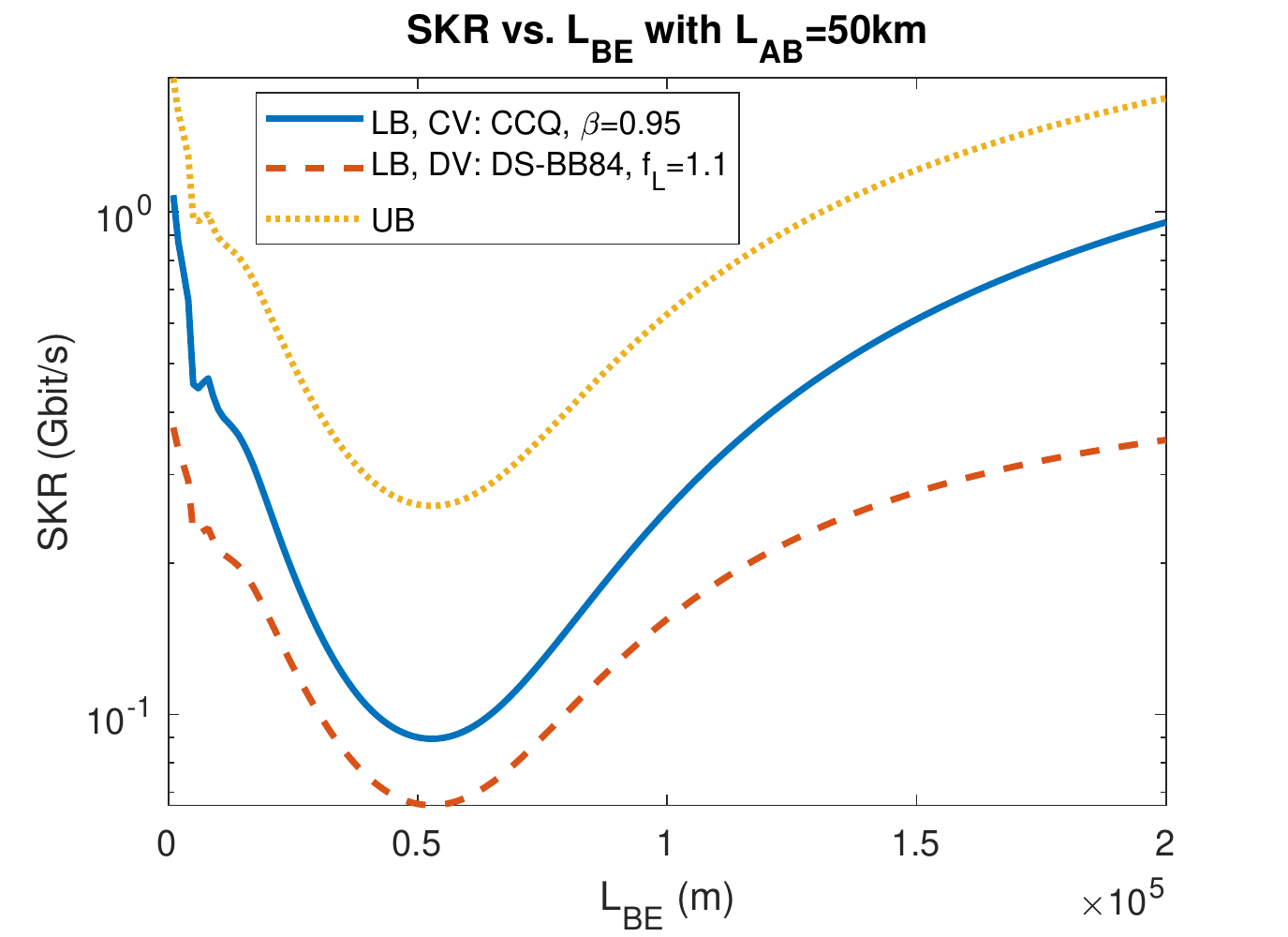}}
\subfigure[]{
\label{Index202012141329_2_50km_D0}
\includegraphics[width=0.45\textwidth]{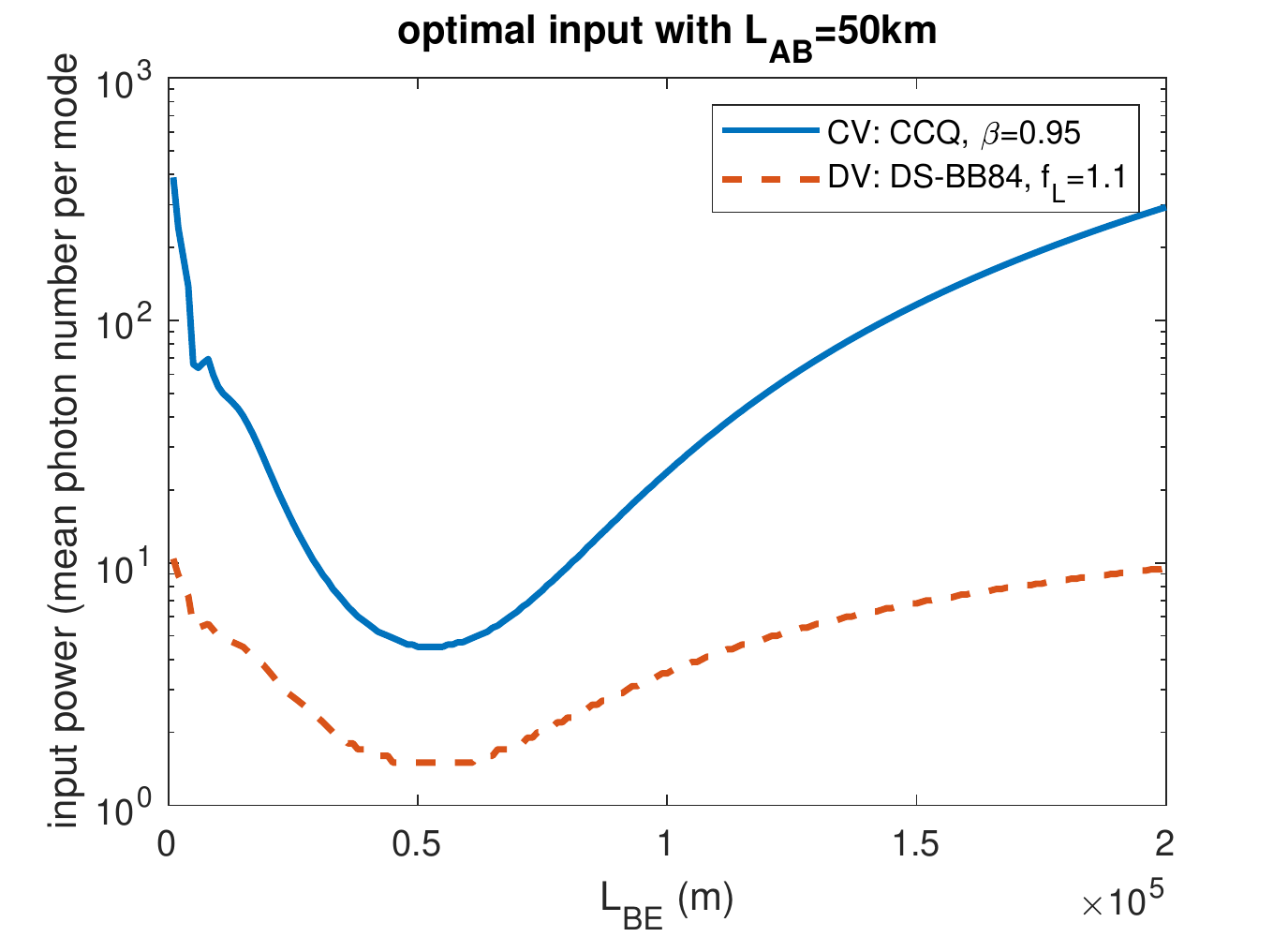}}
\caption{(a) SKR lower bound (LB) for Gaussian modulated CV-QKD and DS-BB84 versus $L_{BE}$ with optimized input power. Upper bound (UB) is also included for comparison. Reconciliation efficiencies are set to $\beta=0.95, f_L=1.1$. $L_{AB}=50$km.
  Transmission center wavelength $\lambda$ is set to 1550nm. Transmitted Gaussian beam waist radius is set to $W_0=r_a=10$cm. $r_b=r_e=10$cm. R=1Gbit/s. (b) Corresponding optimal input power for Gaussian modulated CV-QKD and DS-BB84 in Fig.~\ref{Index202012141329_1_50km_D0}.}
%\label{Fig.main}
\end{figure}

In Fig.~\ref{Index202012141329_1_50km_D0} we apply Gaussian-modulated CV-QKD  (with coherent states, heterodyne detection and reverse reconciliation) and Decoy-state BB84 (DS-BB84) protocols to our studied scenario. We assume that Alice uses a weak coherent-state source and transmits signal-state pulses  to Bob with $\mu$ mean photons per pulse at $R$ states/s. We use the CCQ (classical-classical-quantum) rate (solid curve) for CV-QKD as in Eq.~(69) from~\cite{pan2019secret} and  we use Eq.~(95) with reconciliation efficiency $f_L$ from~\cite{pan2019secret} for DS-BB84 (dashed curves). Here CCQ means that both Alice and Bob have performed measurements~\cite{pan2019secret}. We also include the upper bound as the dotted curve. The numerically optimized input power $\mu$ is plotted in Fig.~\ref{Index202012141329_2_50km_D0}. We can see that both SKR and the optimal input power reaches their global minimum around the point where $L_{BE}=L_{AB}$ while CV-QKD SKR is higher than DS-BB84. When Eve is not at her optimal eavesdropping distance, the SKR can go higher with the input power also increased.

%plotted with code Index202012141329 in OneNote Research in QKD

\subsection{$D$ optimized}\label{Doptimized}

\begin{figure}[htbp]
\centering
\includegraphics[width=8.8cm]{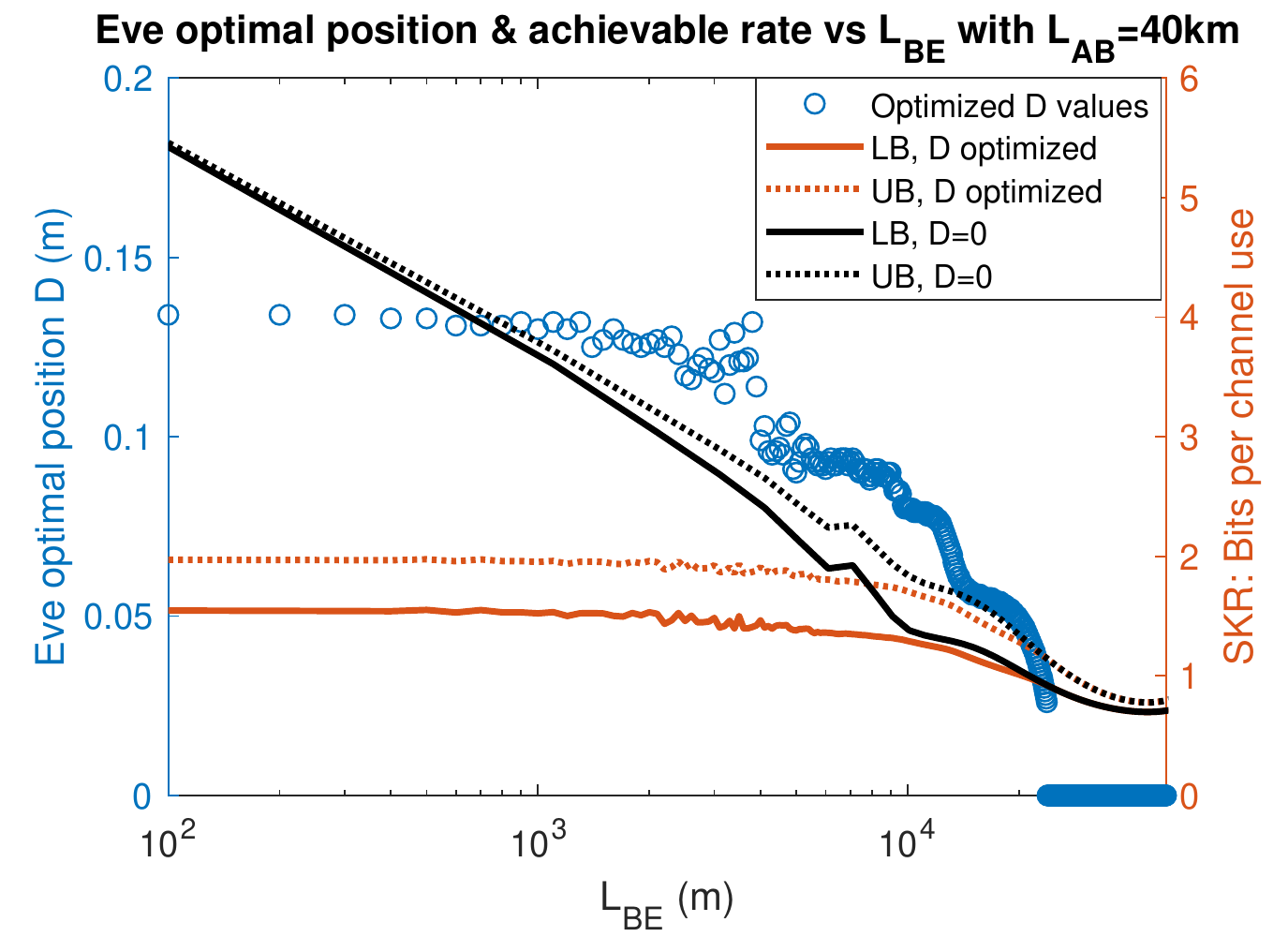}
\caption{Lower bounds (LB) and upper bounds (UB) versus $L_{BE}$ with optimized Eve's position $D$. 
  Transmission center wavelength $\lambda$ is set to 1550nm.  Bob and Eve aperture radius are  set to $r_b=r_e=10$cm. $W_0=10$cm. $L_{AB}=40$km. Reconciliation efficiency $\beta=1$.  $D=0$ case is also included in black curves for comparison. %\textbf{Problem occurred! When including also the D=0 result, the optimized D result is not always lower than D=0 result. Might need to increase precision on optimizing D algorithm, wait for Index202012012155 data to see what to do. Problem solved: two curves had different distance range but was plotted with the same distance range.} %Here the wavelength is set to 1550nm. %the radius of Alice and Bob apertures are set to $5$\text{cm}. 
 \label{Index202012021759_2_AB40km}}
%\end{minipage}
\end{figure}

In this section we investigate the case where Eve can optimize her position, namely she can optimize $D$ as $L_{BE}$ changes. We set $\beta=1$ and take input power $\mu$ to infinity as that is the optimal input power in this case. In Fig.~\ref{Index202012021759_2_AB40km} we set $L_{AB}=40km$ and numerically optimize Eve's position $D$ with different values of $L_{BE}$. Here the achievable rate lower and upper bounds are in red solid and red dotted curves while the optimal position value of Eve $D$ is marked with blue circles. The achievable rate lower and upper bounds with $D=0$ is also included as the black solid and black dotted curves for comparison.

%ploted by Index202012021759 in Onenote 
Here we can see that when $L_{BE}$ is small, Eve will be able to collect more power by optimizing her position $D$, which decreases the achievable rate compared with the $D=0$ case. However, when $L_{BE}$ is larger that the reconverging process of the transmitted beam is complete, Eve's optimal position remains zero so she would not be able to gain any advantage with position optimization beyond that point. Moreover, when $L_{BE}$ is small, Eve's optimal position $D$ is around 14cm which is near the edge of the cropped Gaussian beam since more photons can be collected here before the beam is fully reconverged. When $L_{BE}$ increases, Eve's optimal position $D$ shows a decreasing trend, which is also due to the reconverging process, as more and more power starts to focus at the center. %The optimal position $D$ already converges to zero a little before the optimal eavesdropping distance $L_{BE}=L_{AB}$ and remain that way

\begin{figure}[htbp]
\centering
\includegraphics[width=8.8cm]{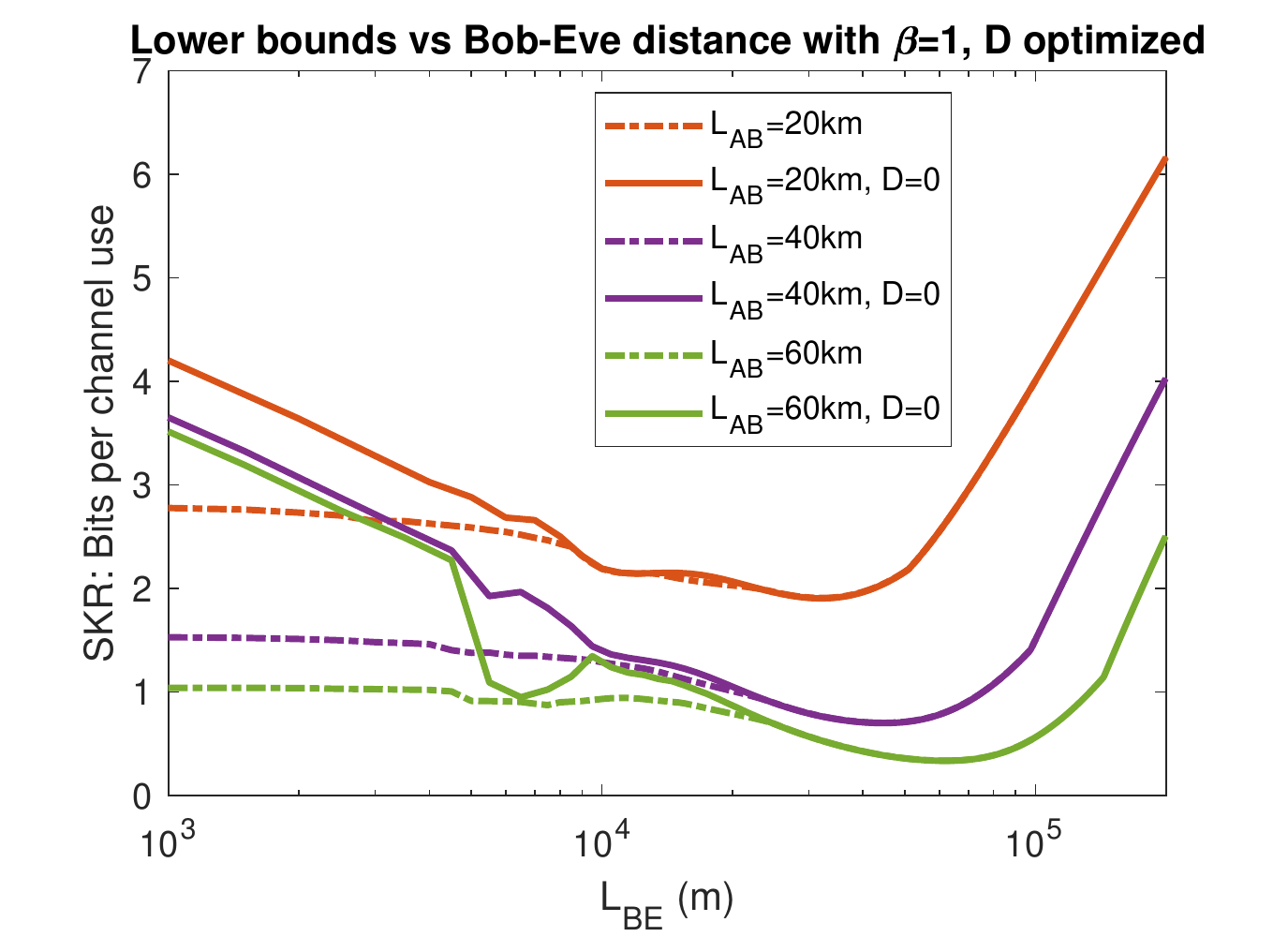}
\caption{Lower bounds versus $L_{BE}$ with optimized Eve's position $D$. 
  Transmission center wavelength $\lambda$ is set to 1550nm.  Bob and Eve aperture radius are  set to $r_b=r_e=10$cm. $W_0=10$cm. $L_{AB}=40$km.  $D=0$ case is also included in solid curves for comparison. %\textbf{Problem occurred! When including also the D=0 result, the optimized D result is not always lower than D=0 result. Might need to increase precision on optimizing D algorithm, wait for Index202012012155 data to see what to do. Problem solved: two curves had different distance range but was plotted with the same distance range.} %Here the wavelength is set to 1550nm. %the radius of Alice and Bob apertures are set to $5$\text{cm}. 
 \label{Index202012111456}}
%\end{minipage}
\end{figure}
%ploted by Index202012111456 in Onenote 

In Fig.~\ref{Index202012111456} we compare the lower bounds of the achievable rate with different $L_{AB}$ and different eavesdropping strategies. For the solid curves $D=0$ and for the dot-dashed curves $D$ is numerically optimized. It is shown that by optimizing her position $D$ Eve can suppress the achievable rate lower bounds when $L_{BE}$ is small with the curves more flattened out. However we can see that Eve still cannot exceed her advantage over Bob when she is at the optimal eavesdropping distance $L_{BE}^\text{optimal}\approx L_{AB}$.

\begin{figure}[htbp]
\centering  %图片全局居中
\subfigure[]{
\label{Index202012141759_1_50km_Doptimal}
\includegraphics[width=0.45\textwidth]{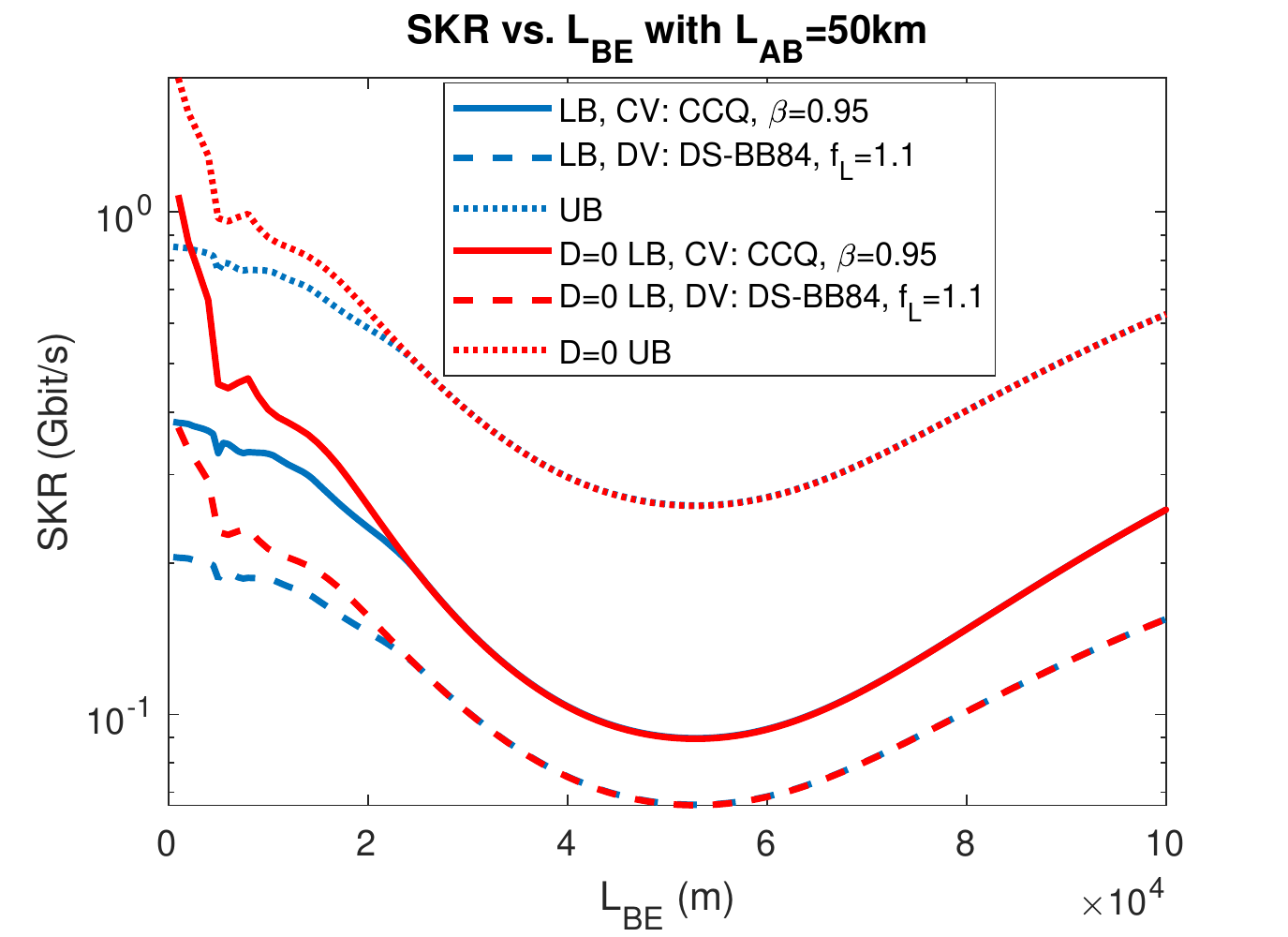}}
\subfigure[]{
\label{Index202012141759_2_50km_Doptimal}
\includegraphics[width=0.45\textwidth]{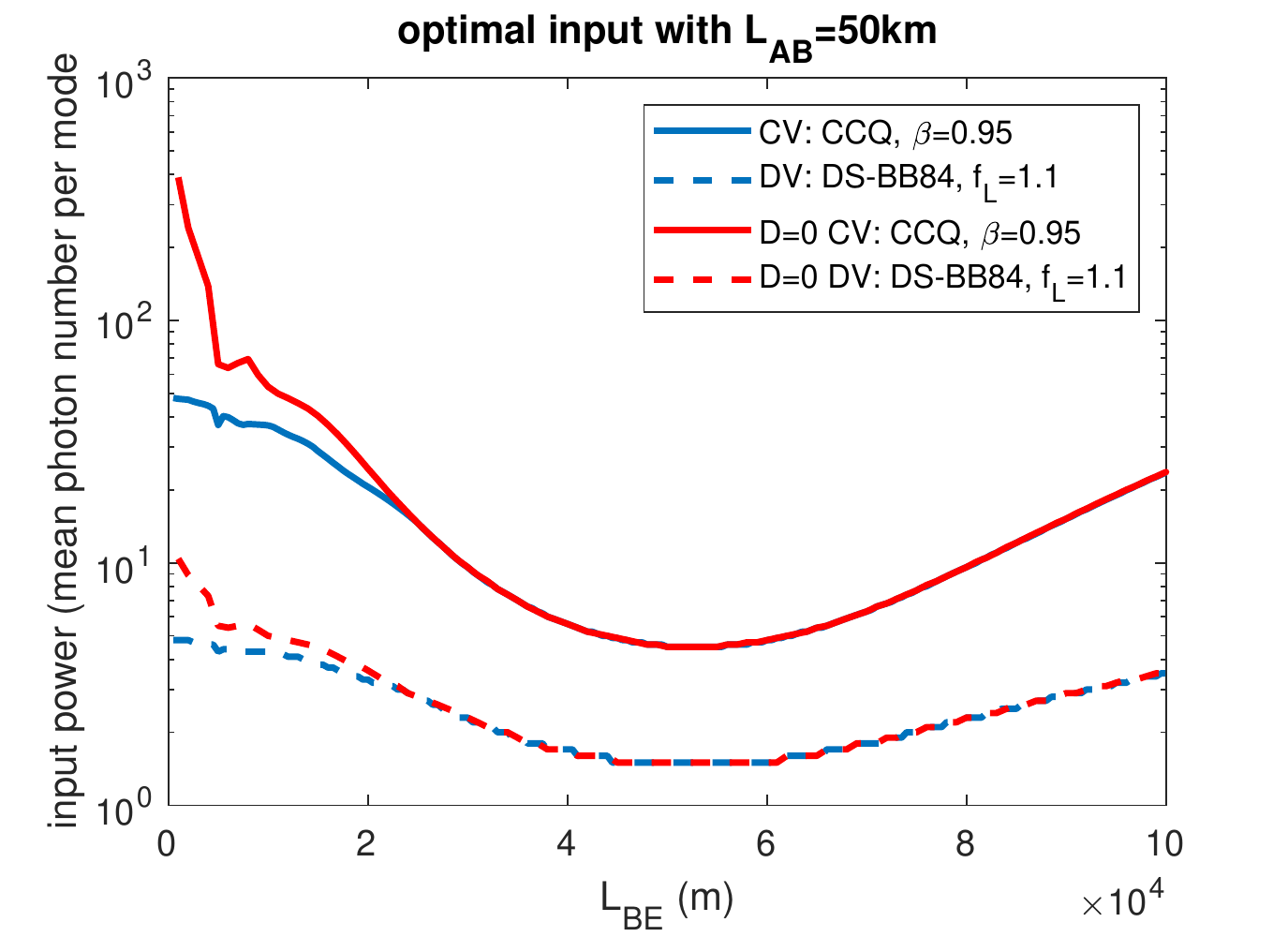}}
\caption{(a) SKR lower bound (LB) for Gaussian modulated CV-QKD and DS-BB84 versus $L_{BE}$ with optimized input power. Upper bound (UB) is also included for comparison. Reconciliation efficiencies are set to $\beta=0.95, f_L=1.1$. $L_{AB}=50$km.
  Transmission center wavelength $\lambda$ is set to 1550nm. Transmitted Gaussian beam waist radius is set to $W_0=r_a=10$cm. $r_b=r_e=10$cm. R=1Gbit/s. (b) Corresponding optimal input power for Gaussian modulated CV-QKD and DS-BB84 in Fig.~\ref{Index202012141759_1_50km_Doptimal}. Curves from Fig.~\ref{Index202012141329_1_50km_D0} and \ref{Index202012141329_2_50km_D0} are also retained for comparison.}
%\label{Fig.main}
\end{figure}

In Fig.~\ref{Index202012141759_1_50km_Doptimal} we apply Gaussian-modulated CV-QKD  (with coherent states, heterodyne detection and reverse reconciliation) and DS-BB84 protocols when Eve's position $D$ is optimized. The numerically optimized input power $\mu$ is plotted in Fig.~\ref{Index202012141759_2_50km_Doptimal}. We also retain the curves from $D=0$ case   for comparison. We can see that both SKR and the optimal input power decreases compared to the $D=0$ case when $L_{BE}$ is small due to the advantages gained by Eve with position optimization.

%plotted with code Index202012142309 in OneNote Research in QKD

\section{Eve before Bob}\label{BeforeBob}
\begin{figure}[htbp]%[htbp] %% Figure 4 
\centering{\includegraphics[height=13pc]{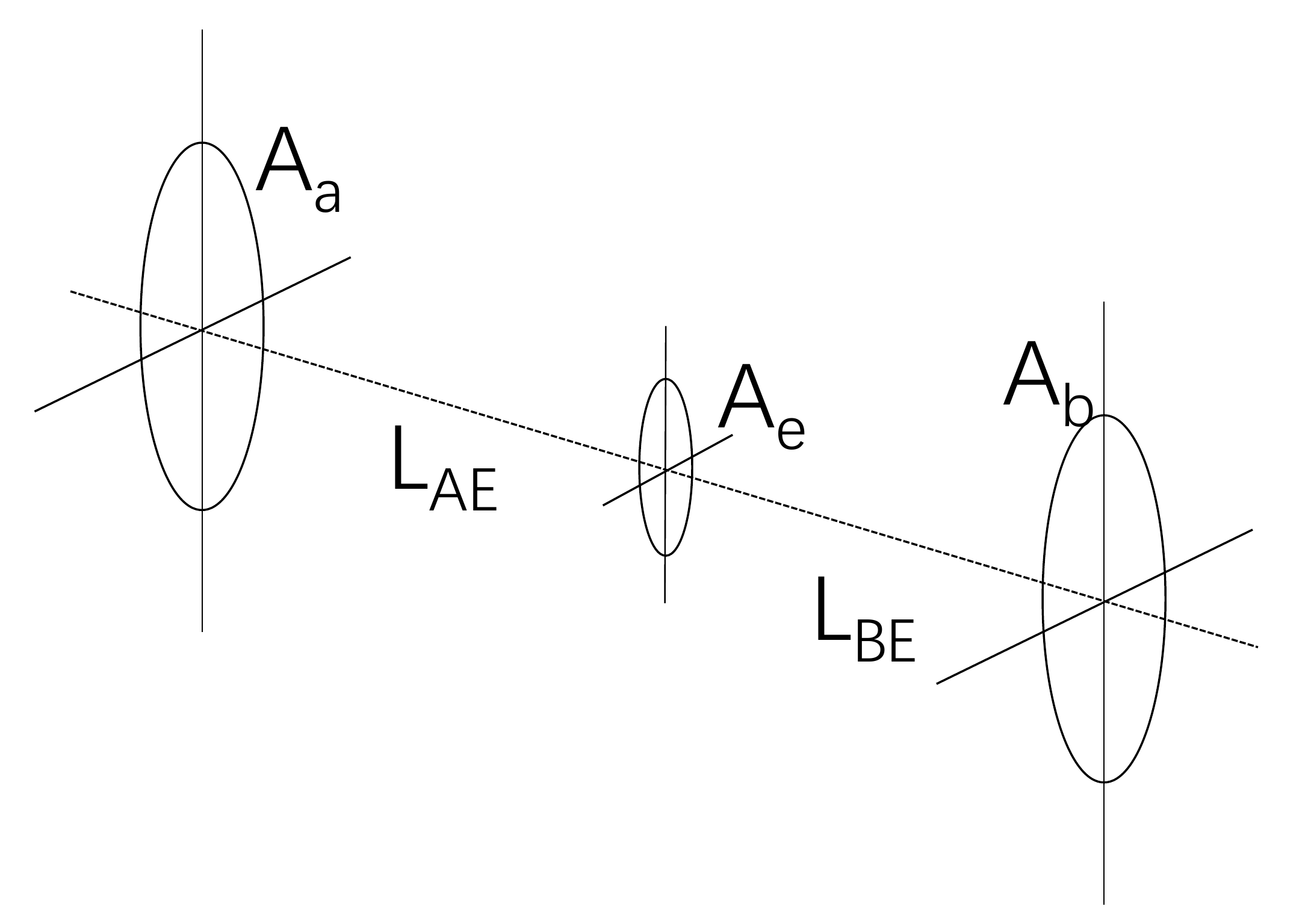}}
\caption{Geometric setup when the eavesdropper Eve is before Bob. 
$L_{AE}$ is the   distance between Alice and Eve and $L_{BE}$ is the distance between Eve and Bob. It is quite straightforward to see that Eve would be able to  collect the most of photons by placing her aperture on the beam transmission axis.%Eve can collect photons outside of the exclusion zone.
\label{EvebeforeBob}}
\end{figure}
\begin{figure}[htbp]
\centering
\includegraphics[width=8.8cm]{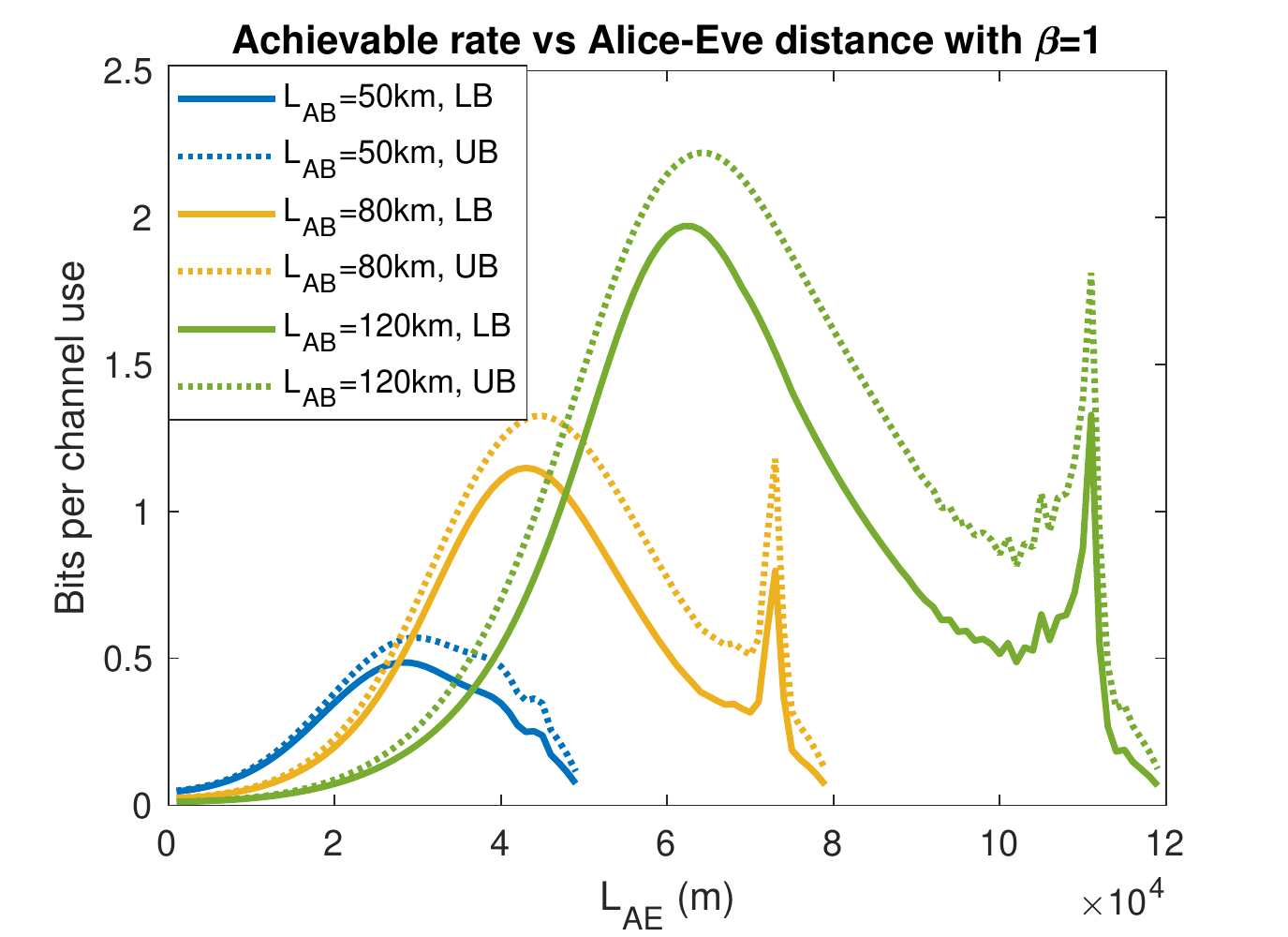}
\caption{Lower bounds (LB) and upper bounds (UB) versus $L_{AE}$ with Eve before Bob. 
  Transmission center wavelength $\lambda$ is set to 1550nm.  Bob and Eve aperture radius are  set to $r_b=r_e=10$cm. $W_0=r_a=10$cm. Different $L_{AB}$ is shown in the legend. %\textbf{Problem occurred! When including also the D=0 result, the optimized D result is not always lower than D=0 result. Might need to increase precision on optimizing D algorithm, wait for Index202012012155 data to see what to do. Problem solved: two curves had different distance range but was plotted with the same distance range.} %Here the wavelength is set to 1550nm. %the radius of Alice and Bob apertures are set to $5$\text{cm}. 
 \label{Index202012131500}}
%\end{minipage}
\end{figure}

%As is in Fig.~\ref{EvePo2}, 
\noindent As in Fig.~\ref{EvebeforeBob}, in this section we investigate the case where Eve is before Bob. In this case, Eve would be able to collect the most power when she place her aperture on the beam transmission axis. In Fig.~\ref{Index202012131500} we plot the lower bounds (LB) and upper bounds (UB) as functions of $L_{AE}$ with different $L_{AB}$. We can see that the achievable rate reaches its peak (global maximum) near the point where $L_{AE}=L_{BE}$, which corroborates with what we observed before. This suggests that Eve's strategy would be getting close to either Alice or Bob to gain advantages over the communication parties, which further suggests that an exclusion zone~\cite{pan2020secretExZo} set by the communication parties would serve as a good defense strategy in this case. We also notice that when Eve is too close to Bob there would be some oscillations due to the constructive and destructive interference on Bob's receiving aperture that leads to  local maximums of the achievable rate. This suggests that although generally Eve will be able to gain advantages by placing her aperture either closer to Alice or closer to Bob, it would be safer for her to get closer to Alice than to Bob to avoid the potential local maximum peaks.

%ploted by Index202012131500 in Onenote 

\begin{figure}[htbp]
\centering
\includegraphics[width=10cm]{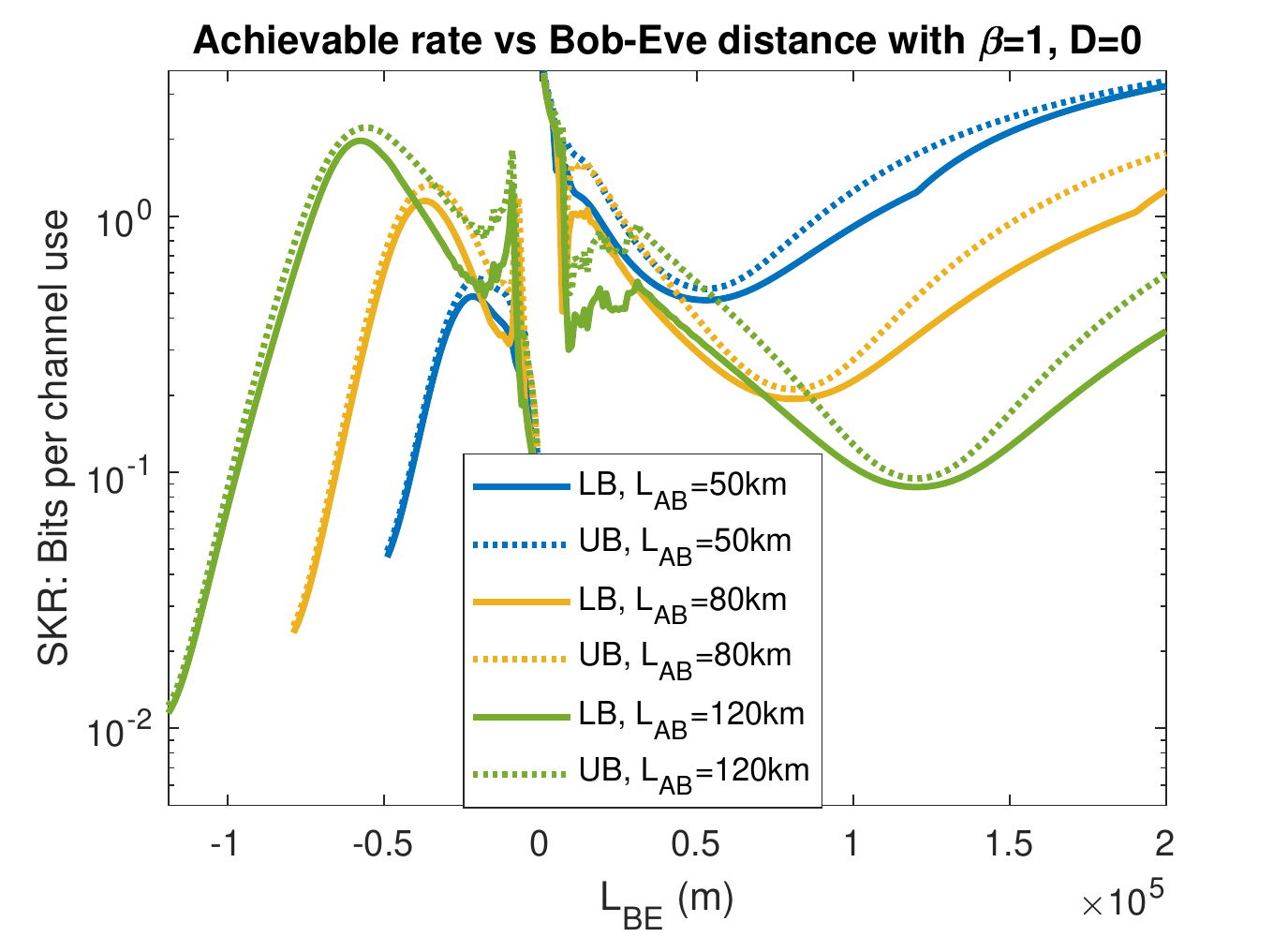}
\caption{Lower bounds (LB) and upper bounds (UB) versus $L_{BE}$. Here $L_{BE}$ is negative if  Eve is before Bob and positive if Eve is behind Bob ($D=0$).
  Transmission center wavelength $\lambda$ is set to 1550nm.  Bob and Eve aperture radius are  set to $r_b=r_e=10$cm. $W_0=r_a=10$cm. Different $L_{AB}$ is shown in the legend. %\textbf{Problem occurred! When including also the D=0 result, the optimized D result is not always lower than D=0 result. Might need to increase precision on optimizing D algorithm, wait for Index202012012155 data to see what to do. Problem solved: two curves had different distance range but was plotted with the same distance range.} %Here the wavelength is set to 1550nm. %the radius of Alice and Bob apertures are set to $5$\text{cm}. 
 \label{Index202012231126}}
%\end{minipage}
\end{figure}

To have a better understanding of Eve's strategy through comparison, in Fig.~\ref{Index202012231126} we set $L_{BE}$ to negative when Eve is before Bob and positive if Eve is behind Bob and plot the achievable rate lower and upper bounds as functions of $L_{BE}$ with different $L_{AB}$ specified in the legend, combining the Eve before Bob case and the Eve behind Bob case. Here for the Eve behind Bob case, we set $D=0$. We can see that when $L_{AB}$ is small, Eve is able to eavesdrop more with her aperture before Bob even if she is not close to either Alice or Bob since when $L_{AB}$ is small, the beam divergence  is not significant so Eve would be able to collect more photons in-between the transmission channel from Alice to Bob than behind Bob. However, when $L_{AB}$ increases, although Eve can always gain advantages by getting very close to the communication parties, in reality because of the risk of being found, Eve would benefit more by placing her aperture behind Bob and move to the point $L_{BE}^\text{optimal}$, as the cropped beam would reconverge with a large portion of its power collectable by Eve and lower chances of being found, like we showed in Sec.~\ref{LowerBoundswithOptimizedInputPower}. From Alice's side, a shorter transmission distance is always going to be safer, especially if combined with an exclusion zone set by the communication parties to prevent Eve from approaching either Alice or Bob. In a long-distance transmission scenario, aside from an exclusion zone near the communication parties' apertures, similar measures around the point where $L_{BE}=L_{AB}$ is also crucial for safety issues although this might not be very realistic in implementations. Another alternative would be to prevent the cropped Gaussian beam from propagating farther, eg, with a light barrier behind Bob.

%ploted by Index202012231126 in Onenote 

%Clearly, from Fig.~\ref{Index202007281001}  we can see that the SKR lower bounds converge to non-zero values either when transmission distance $L$ is sufficiently small or large. When $L$ is small, it's easy to see that converging trend as $L$ simply disappears from our Eqs.~(\ref{PBob}) and (\ref{PEve}) for $P_{Bob}$ and $P_{Eve}$ when we take $L\rightarrow0$, which thus give us distance independent rates. This can be better understood through the structure of Gaussian beam as in Fig.~\ref{GaussianBeam}. We can see that the beam width decreases to the waist radius as approaching the focus of the beam ($z=0$). We can also see that as we approach the focus of the beam the beam width decreases more and more slowly to converge to the waist radius $W_0$. This leads to the non-zero rate when $L$ is small. %for certain choices of parameters we might be able to have a nearly constant rate that doesn't change with increasing transmission distance. %(We can consider including the more specific explanation for this threshold for later followup paper since this one is a little hard to derive) 
%In Fig.~\ref{geo1} we plot the SKR lower bound versus transmission distance at wavelength $\lambda=0.5 \text{mm}$,  $r_a=r_b=5\text{cm}$. 

\begin{figure}[htbp]
\centering  %图片全局居中
\subfigure[]{
\label{Index202012141814_1_80km_EvebeforeBob}
\includegraphics[width=0.45\textwidth]{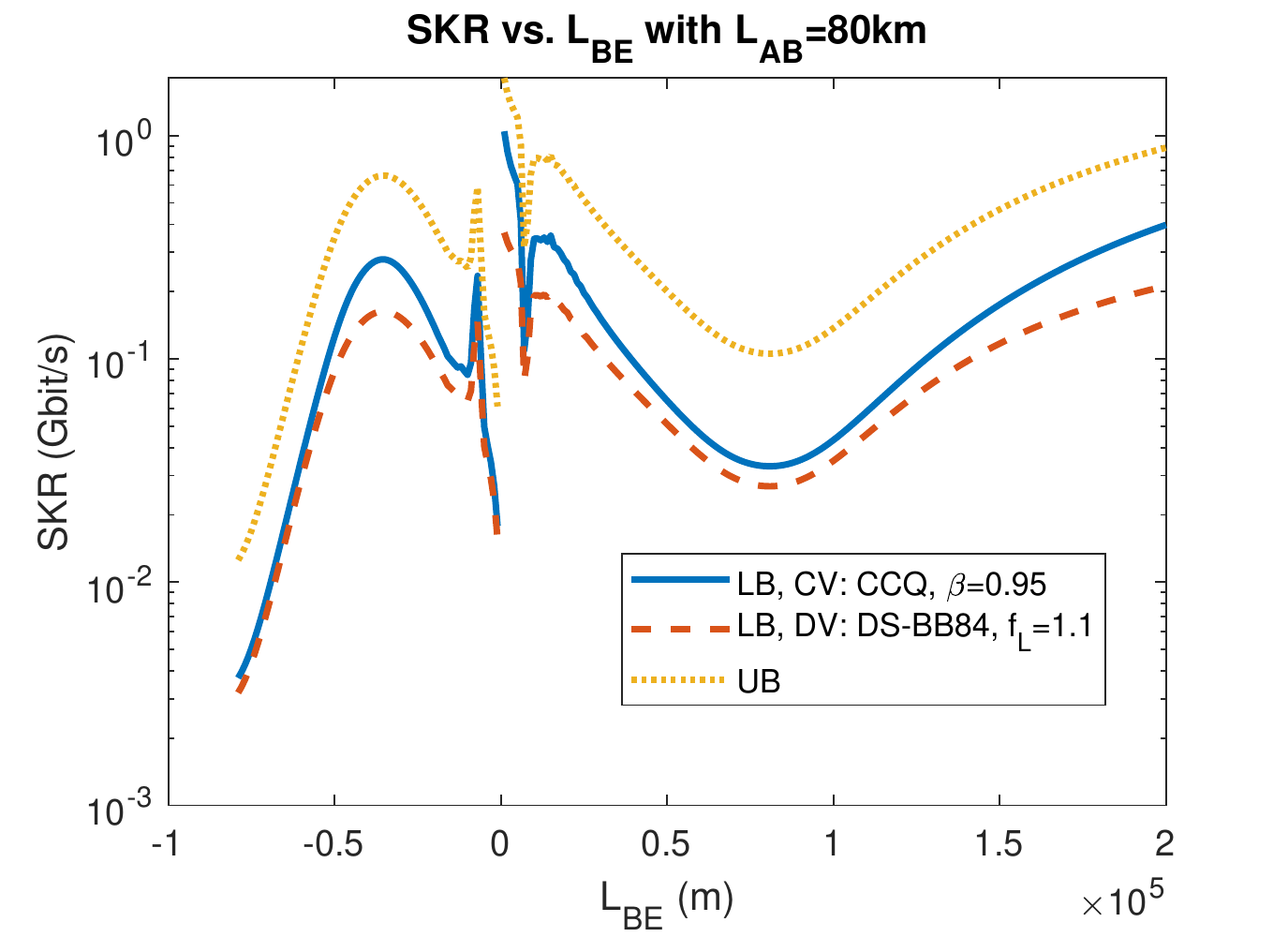}}
\subfigure[]{
\label{Index202012141814_2_80km_EvebeforeBob}
\includegraphics[width=0.45\textwidth]{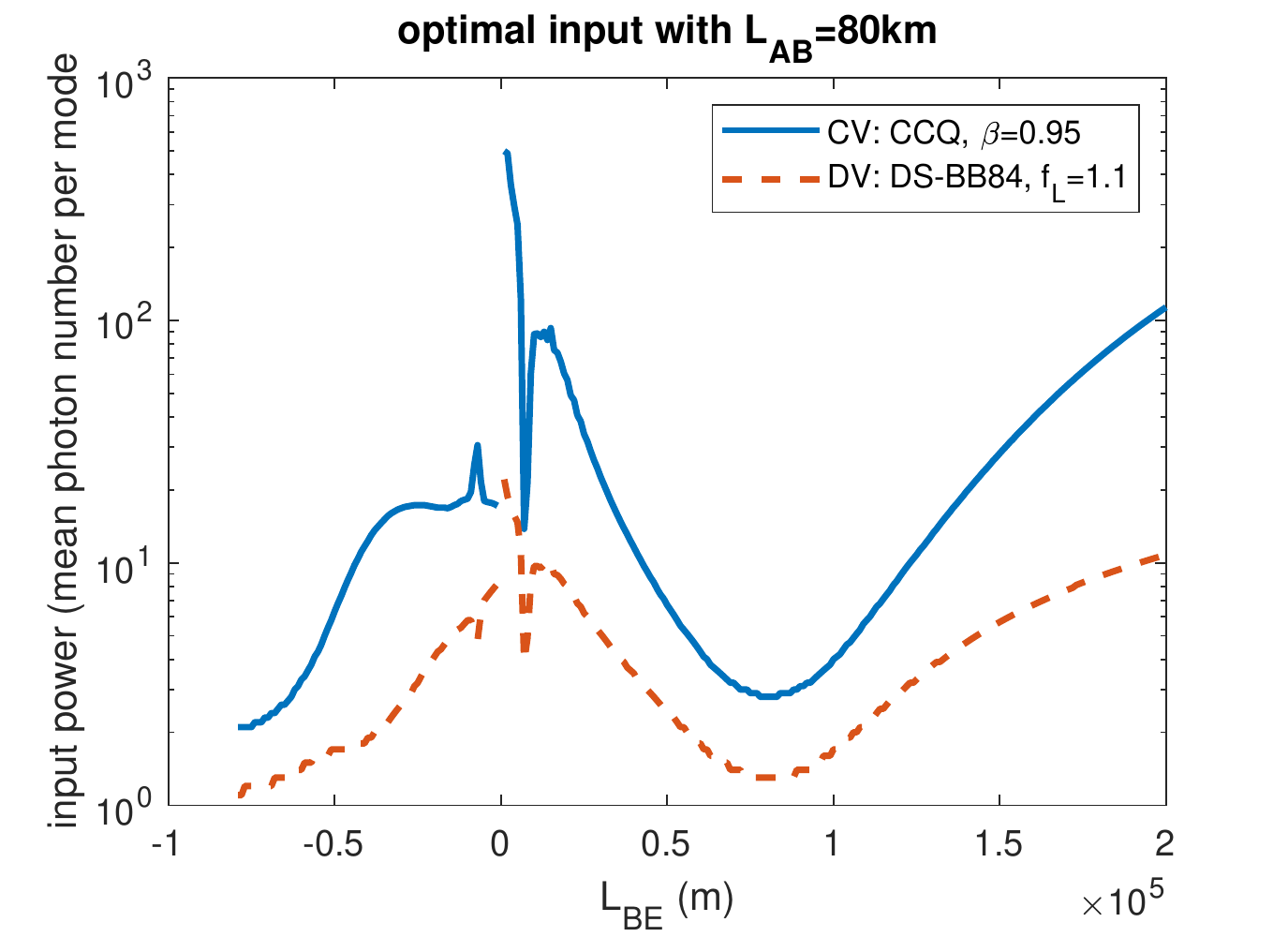}}
\caption{(a) SKR lower bound (LB) for Gaussian modulated CV-QKD and DS-BB84 versus $L_{BE}$ with optimized input power. Upper bound (UB) is also included for comparison. Reconciliation efficiencies are set to $\beta=0.95, f_L=1.1$. $L_{AB}=80$km.
  Transmission center wavelength $\lambda$ is set to 1550nm. Transmitted Gaussian beam waist radius is set to $W_0=r_a=10$cm. $r_b=r_e=10$cm. R=1Gbit/s. (b) Corresponding optimal input power for Gaussian modulated CV-QKD and DS-BB84 in Fig.~\ref{Index202012141814_1_80km_EvebeforeBob}. Here a negative $L_{BE}$ means that Eve is before Bob while a positive $L_{BE}$ means Eve is behind Bob ($D=0$).}
%\label{Fig.main}
\end{figure}
%plotted with code Index202012141814 in OneNote Research in QKD

In Fig.~\ref{Index202012141814_1_80km_EvebeforeBob} we apply Gaussian-modulated CV-QKD  (with coherent states, heterodyne detection and reverse reconciliation) and DS-BB84 protocols when Eve is before Bob. %We assume that Alice uses a weak coherent-state source and transmits signal-state pulses  to Bob with $\mu$ mean photons per pulse at $R$ states/s. We use the CCQ (classical-classical-quantum) rate (solid curve) for CV-QKD as in Eq.~(69) from~\cite{pan2019secret} and  we use Eq.~(95) with reconciliation efficiency $f_L$ from~\cite{pan2019secret} for DS-BB84 (dashed curves). Here CCQ indicates that both Alice and Bob have performed measurements~\cite{pan2019secret}. We also include the upper bound as  dotted curves. 
The numerically optimized input power $\mu$ is plotted in Fig.~\ref{Index202012141814_2_80km_EvebeforeBob}. Here we also use negative $L_{BE}$ to denote that Eve is before Bob and positive $L_{BE}$ to denote that Eve is behind Bob ($D=0$). We can see that when Eve is before Bob, the SKR reaches its global maximum peak near the point where $L_{BE}=L_{AB}/2$ and there is a local maximum peak when Eve is close to Bob.

\section{Summary}

%\noindent We have analyzed SKR lower bounds for realistic scenarios over FSO satellite-to-satellite channel where Eve can optimize her position to gain an advantage when she is close to Bob. This actually suggests that simple measures such as setting an exclusion zone around Bob's receiver could be very effective %should be taken to keep Eve away from Bob up to a certain distance depending on the channel parameters 
%to ensure higher security.

\noindent In this paper, we have analyzed satellite-to-satellite secret key distillation with Eve having a dynamically-positioned limited-sized aperture by calculating the  power reception of Bob and Eve respectively. % achievable  key rate lower bounds with respect to relevant channel parameters for a  realistic scenario in FSO satellite-to-satellite secret key distillation where Eve has a dynamic-positioned limited-sized aperture. 
For the case where Eve is behind Bob, we show that Eve's optimal eavesdropping distance to Bob $L_{BE}^\text{optimal}$ in a long-distance transmission scenario ($L_{AB}$ is large) is approximately equal to Alice-to-Bob distance $L_{AB}$. When $L_{AB}$ is small we show that the achievable  key rate lower bounds become close to the results from the study of Arago Spot since the Gaussian beam transmitted can be viewed as approximately collimated. We further showed that when Eve can move her aperture off the beam transmission axis she can gain advantages when approaching Bob but still cannot exceed the $L_{BE}^\text{optimal}$ case. When Eve is before Bob, we showed that the achievable  key rate would be the highest if Eve is near the middle of Alice and Bob, and Eve's strategy would be to approach Alice or Bob, but avoid the local maximums caused by constructive  interference on Bob's receiving aperture when close to Bob. This also suggests that an exclusion zone would be a very effective defense strategy in this case. We also compared the case with Eve before Bob and Eve behind Bob to analyze Eve's strategy with respect to different $L_{AB}$ and the communication parties' possible corresponding measures. Finally, for these studied scenarios we  applied our calculations on Gaussian-modulated CV-QKD and DS-BB84 protocols SKR bounds and showed similar phenomena with comparison between them.

\section*{Funding}
This paper was supported in part by L3Harris and NSF.

\section*{Acknowledgement}
 Z. P. thankfully acknowledges helpful discussions with Saikat Guha, Kaushik P. Seshadreesan and John Gariano from the University of Arizona, Jeffrey H. Shapiro from Massachusetts Institute of Technology and William Clark, Mark R. Adcock from General Dynamics.

%\end{comment}

\section*{Disclosures}

\noindent The authors declare no conflicts of interest.

%%%%%%%%%% If using BibTeX:
\bibliography{sample}

\begin{thebibliography}{10}
\newcommand{\enquote}[1]{``#1''}

\bibitem{bennett1984quantum}
C.~H. Bennett and G.~Brassard, \enquote{Quantum cryptography: Public key
  distribution and coin tossing,} in \emph{Proceedings of IEEE International
  Conference on Computers, Systems and Signal Processing,}  (New York, 1984),
  pp. 175--179.

\bibitem{inoue2002differential}
K.~Inoue, E.~Waks, and Y.~Yamamoto, \enquote{Differential phase shift quantum
  key distribution,} {\protect\JournalTitle{Physical review letters}}
  \textbf{89}, 037902 (2002).

\bibitem{hwang2003quantum}
W.-Y. Hwang, \enquote{Quantum key distribution with high loss: toward global
  secure communication,} {\protect\JournalTitle{Physical Review Letters}}
  \textbf{91}, 057901 (2003).

\bibitem{pan2017quantum}
Z.~Pan, J.~Cai, and C.~Wang, \enquote{Quantum key distribution with high order
  fibonacci-like orbital angular momentum states,}
  {\protect\JournalTitle{International Journal of Theoretical Physics}}
  \textbf{56}, 2622--2634 (2017).

\bibitem{PhysRevLett.108.130503}
H.-K. Lo, M.~Curty, and B.~Qi, \enquote{Measurement-device-independent quantum
  key distribution,} {\protect\JournalTitle{Phys. Rev. Lett.}} \textbf{108},
  130503 (2012).

\bibitem{PhysRevLett.108.130502}
S.~L. Braunstein and S.~Pirandola, \enquote{Side-channel-free quantum key
  distribution,} {\protect\JournalTitle{Phys. Rev. Lett.}} \textbf{108}, 130502
  (2012).

\bibitem{PhysRevA.86.062319}
X.~Ma and M.~Razavi, \enquote{Alternative schemes for
  measurement-device-independent quantum key distribution,}
  {\protect\JournalTitle{Phys. Rev. A}} \textbf{86}, 062319 (2012).

\bibitem{LPFPP18}
F.~Laudenbach, C.~Pacher, C.-H.~F. Fung, A.~Poppe, M.~Peev, B.~Schrenk,
  M.~Hentschel, P.~Walther, and H.~H{\"u}bel, \enquote{Continuous-variable
  quantum key distribution with gaussian modulation—the theory of practical
  implementations,} {\protect\JournalTitle{Advanced Quantum Technologies}}
  \textbf{1}, 1800011 (2018).

\bibitem{DL15}
E.~Diamanti and A.~Leverrier, \enquote{Distributing secret keys with quantum
  continuous variables: Principle, security and implementations,}
  {\protect\JournalTitle{Entropy}} \textbf{17}, 6072--6092 (2015).

\bibitem{zhang2019continuous}
Y.~Zhang, Z.~Li, Z.~Chen, C.~Weedbrook, Y.~Zhao, X.~Wang, Y.~Huang, C.~Xu,
  X.~Zhang, Z.~Wang \emph{et~al.}, \enquote{Continuous-variable qkd over 50 km
  commercial fiber,} {\protect\JournalTitle{Quantum Science and Technology}}
  \textbf{4}, 035006 (2019).

\bibitem{zhang2020long}
Y.-C. Zhang, Z.~Chen, S.~Pirandola, X.~Wang, C.~Zhou, B.~Chu, Y.~Zhao, B.~Xu,
  S.~Yu, and H.~Guo, \enquote{Long-distance continuous-variable quantum key
  distribution over 202.81 km fiber,} {\protect\JournalTitle{arXiv preprint
  arXiv:2001.02555}}  (2020).

\bibitem{pan2019secret}
Z.~Pan, K.~P. Seshadreesan, W.~Clark, M.~R. Adcock, I.~B. Djordjevic, J.~H.
  Shapiro, and S.~Guha, \enquote{Secret key distillation across a quantum
  wiretap channel under restricted eavesdropping,} {\protect\JournalTitle{arXiv
  preprint arXiv:1903.03136}}  (2019).

\bibitem{8849223}
Z.~Pan, K.~P. {Seshadreesan}, W.~{Clark}, M.~R. {Adcock}, I.~B. {Djordjevic},
  J.~H. {Shapiro}, and S.~{Guha}, \enquote{Secret key distillation over a pure
  loss quantum wiretap channel under restricted eavesdropping,} in \emph{2019
  IEEE International Symposium on Information Theory (ISIT),}  (2019), pp.
  3032--3036.

\bibitem{pan2020secretOE}
Z.~Pan and I.~B. Djordjevic, \enquote{Secret key distillation over
  satellite-to-satellite free-space optics channel with a limited-sized
  aperture eavesdropper in the same plane of the legitimate receiver,}
  {\protect\JournalTitle{Optics Express}} \textbf{28}, 37129--37148 (2020).

\bibitem{pan2020secretQ2}
Z.~Pan, J.~Gariano, W.~Clark, and I.~B. Djordjevic, \enquote{Secret key
  distillation over realistic satellite-to-satellite free-space channel,} in
  \emph{Quantum 2.0,}  (Optical Society of America, 2020), pp. QTh7B--15.

\bibitem{pan2020secretExZo}
Z.~Pan and I.~B. Djordjevic, \enquote{Secret key distillation over realistic
  satellite-to-satellite free-space channel: exclusion zone analysis,}
  {\protect\JournalTitle{arXiv preprint arXiv:2009.05929}}  (2020).

\bibitem{pan2020securityICTON}
Z.~Pan and I.~B. Djordjevic, \enquote{Security of satellite-based cv-qkd under
  realistic assumptions,} in \emph{2020 22nd International Conference on
  Transparent Optical Networks (ICTON),}  (IEEE, 2020), pp. 1--4.

\bibitem{pan2020secretSPPCom}
Z.~Pan, J.~Gariano, and I.~B. Djordjevic, \enquote{Secret key distillation over
  satellite-to-satellite free-space channel with eavesdropper dynamic
  positioning,} in \emph{Signal Processing in Photonic Communications,}
  (Optical Society of America, 2020), pp. SpTu3I--4.

\bibitem{harvey1984spot}
J.~E. Harvey and J.~L. Forgham, \enquote{The spot of arago: new relevance for
  an old phenomenon,} {\protect\JournalTitle{American journal of Physics}}
  \textbf{52}, 243--247 (1984).

\bibitem{reisinger2017relative}
T.~Reisinger, P.~Leufke, H.~Gleiter, and H.~Hahn, \enquote{On the relative
  intensity of poisson’s spot,} {\protect\JournalTitle{New Journal of
  Physics}} \textbf{19}, 033022 (2017).

\bibitem{fischer2007dark}
P.~Fischer, S.~E. Skelton, C.~G. Leburn, C.~T. Streuber, E.~M. Wright, and
  K.~Dholakia, \enquote{The dark spots of arago,} {\protect\JournalTitle{Optics
  express}} \textbf{15}, 11860--11873 (2007).

\end{thebibliography}

%%%%%%%%%% If preparing manually:
% \begin{thebibliography}{1}
% \newcommand{\enquote}[1]{``#1''}

% \bibitem{Zhang:14}
% Y.~Zhang, S.~Qiao, L.~Sun, Q.~W. Shi, W.~Huang, L.~Li, and Z.~Yang,
%   \enquote{Photoinduced active terahertz metamaterials with nanostructured
%   vanadium dioxide film deposited by sol-gel method,}
%   {\protect\JournalTitle{Optics Express}} \textbf{22}, 11070--11078 (2014).

% \bibitem{OSA}
% {Optical Society}, \enquote{{OSA Publishing},}
%   \url{http://www.osapublishing.org}.

% \bibitem{FORSTER2007}
% P.~Forster, V.~Ramaswamy, P.~Artaxo, T.~Bernsten, R.~Betts, D.~Fahey,
%   J.~Haywood, J.~Lean, D.~Lowe, G.~Myhre, J.~Nganga, R.~Prinn, G.~Raga,
%   M.~Schulz, and R.~V. Dorland, \enquote{Changes in atmospheric consituents and
%   in radiative forcing,} in \enquote{Climate Change 2007: The Physical Science
%   Basis. Contribution of Working Group 1 to the Fourth assesment report of
%   Intergovernmental Panel on Climate Change,}  S.~Solomon, D.~Qin, M.~Manning,
%   Z.~Chen, M.~Marquis, K.~B. Averyt, M.~Tignor, and H.~L. Miler, eds.
%   (Cambridge University Press, 2007).

% \end{thebibliography}

\end{document}